\documentclass[a4paper,11pt]{article}
\pdfoutput=1
\usepackage{jcappub}
\usepackage{bm}
\usepackage{soul}
\usepackage{latexsym}
\usepackage{dcolumn}
\usepackage{amsfonts,amssymb,amsmath}
\usepackage{graphicx,epsfig}
\usepackage{psfrag}
\usepackage{subfigure}
\usepackage{rotating}
\usepackage{hyperref}
\usepackage{tikz}
\hypersetup{
	unicode=false,          % non-Latin characters in Acrobat’s bookmarks
	pdftoolbar=true,        % show Acrobat’s toolbar?
	pdfmenubar=true,        % show Acrobat’s menu?
	pdffitwindow=false,     % window fit to page when opened
	pdfstartview={FitH},    % fits the width of the page to the window
	pdftitle={My title},    % title
	pdfauthor={Author},     % author
	pdfsubject={Subject},   % subject of the document
	pdfcreator={Creator},   % creator of the document
	pdfproducer={Producer}, % producer of the document
	pdfkeywords={keyword1} {key2} {key3}, % list of keywords
	pdfnewwindow=true,      % links in new window
	colorlinks=true,       % false: boxed links; true: colored links
	linkcolor=red,          % color of internal links
	citecolor=cyan,        % color of links to bibliography
	filecolor=magenta,      % color of file links
	urlcolor=green,           % color of external links
	linktocpage=true
}

\newcommand{\fatdot}{\raisebox{0.25ex}{\tikz\filldraw[black,x=2pt,y=2pt] (0,0) circle (0.2);}}

\def\beq{\begin{equation}}
\def\efq{\end{equation}}
\def\br{\begin{eqnarray}}
\def\er{\end{eqnarray}}
\def\benu{\begin{enumerate}}
	\def\efnu{\end{enumerate}}
\def\nn{\nonumber}

\def\l{\left}
\def\r{\right}
\def\cR{{\cal R}}
\def\d{{\rm d}}
\def\f{\frac}
\def\nn{\nonumber} 

\def\cR{{\mathcal{R}}}
\def\cS{{\mathcal{S}}}

\def\Mpl{{M_{_{\textup{Pl}}}}}

\makeatletter
\gdef\@fpheader{}
\makeatother

\begin{document}
\title{Generating primordial features at large scales in two field models of inflation}
	\author[1,2,3]{Matteo~Braglia,}
	\author[4,2,3]{Dhiraj~Kumar~Hazra,} 
	\author[5]{L.~Sriramkumar,}
	\author[2,3]{Fabio~Finelli}
	\affiliation[1]{DIFA, Dipartimento di Fisica e Astronomia,\\
		Alma Mater Studiorum, Universit\`a degli Studi di Bologna,\\
		Via Gobetti, 93/2, I-40129 Bologna, Italy}
	\affiliation[2]{INAF/OAS Bologna, Osservatorio di Astrofisica e Scienza dello Spazio, \\
		Area della ricerca CNR-INAF, via Gobetti 101, I-40129 Bologna, Italy}
	\affiliation[3]{INFN, Sezione di Bologna, \\
		via Irnerio 46, I-40126 Bologna, Italy}
		\affiliation[4]{The Institute of Mathematical Sciences, HBNI, CIT Campus, Chennai 600113, India}
	\affiliation[5]{Department of Physics, Indian Institute of Technology Madras, 
		Chennai 600036, India}
	\emailAdd{matteo.braglia2@unibo.it, dhiraj@imsc.res.in,
		sriram@physics.iitm.ac.in, fabio.finelli@inaf.it} 
	% \date{\today}

	\abstract
	{ 
		We investigate the generation of features at large scales in the primordial power spectrum (PPS) when inflation is driven by two scalar fields. In canonical single field models of inflation, these features are often generated due to deviations from the slow-roll regime.
		While deviations from slow-roll can be naturally achieved in two field
		models due to a sharp turn in the trajectory in the field space, 
		features at the largest scales of the types suggested by CMB temperature anisotropies are more difficult to achieve in models involving two canonical scalar fields due to the presence of isocurvature fluctuations. We show instead that a coupling between the kinetic terms of the scalar fields can easily produce such features.
We discuss models whose theoretical predictions are consistent with current observations and highlight the implications of our results.
}
	\maketitle
	
	%%%%%%%%%%%%%%%%%%%%%%%%%%%%%%%%%%%%%%%%%%%%%%%%%%%%%%%%%%%%%%%%%%%%%%%%%%%%%%%
	
	\section{Introduction}
	The measurements of anisotropies in the Cosmic Microwave Background 
	(CMB) point to a primordial power spectrum (PPS) of adiabatic nearly Gaussian scalar fluctuations whose deviation from scale invariance has been accurately measured ~\cite{Akrami:2018odb,Akrami:2019izv}.
	The most effective and compelling paradigm to generate perturbations of 
	such type is slow-roll inflation and, in fact, there exist many inflationary models that are remarkably consistent with the cosmological 
	data~\cite{Akrami:2018odb}. However, it has been shown by different methodologies that deviations from the simple power-law in the PPS which is predicted by slow-roll inflation to leading order, can lead to an improvement in the fit to CMB anisotropies, although not at a statistically significant level~\cite{Akrami:2018odb}. Reconstructions of primordial power spectrum directly from CMB data have been indicating features in several publications~\cite{Hannestad2001,Tegmark2002,Mukherjee2003,TocchiniValentini:2005ja,Shafieloo2004,Kogo2005,Leach:2005av,Shafieloo2008,Paykari2010,Nicholson:2009pi,Gauthier2012,Hlozek2012,Vzquez2012,Hazra2013,Hunt2014,Hazra:2014jwa,Brando:2020yvo}.
	
	These features can be broadly divided into the following three types: 
	a lack of power at large scales~\cite{Starobinsky:1992ts,Contaldi:2003zv,Sriramkumar:2004pj,Allahverdi:2006wt,Jain:2007au,Jain:2008dw,Jain:2009pm,Ramirez:2011kk,Ramirez:2012gt,Hazra:2014goa,Hazra2014b,Bousso:2014jca,Hazra:2016fkm,Gruppuso:2015xqa,Hazra:coreforecast,Hergt:2018ksk,Ballardini:2018noo,Ragavendra:2019mek,Ragavendra:2020old}, 
	a dip and burst of oscillations around multipoles $\ell=20$--$40$~\cite{Adams:2001vc,Chen:2006xjb,Covi:2006ci,Hazra:2010ve,Miranda:2012rm,Benetti:2013cja,Romano:2014kla,Chluba:2015bqa,Ballardini:2017qwq}, and smaller oscillations that persist from intermediate to the smallest scales~\cite{Martin:2003sg,Ashoorioon:2006wc,Chen:2008wn,Biswas:2010si,Flauger:2009ab,Pahud:2008ae,Achucarro:2010jv,Aich:2011qv,Hazra:2012vs,Peiris:2013opa,Meerburg:2013dla,Easther:2013kla,Chen:2014cwa,Motohashi:2015hpa,Miranda:2015cea,Ballardini:2016hpi,Ballardini:2019tuc}.

	In the simplest inflationary models involving a single, canonical
	scalar field, features in the power spectrum are generated by a 
	departure from the slow-roll regime or from Bunch-Davies initial conditions.
	The departures from the slow-roll regime, in turn, are often achieved by specific features in the inflaton potential, such as a point of inflection, 
	a step or oscillatory terms. 
	There has been a constant effort in the literature to construct 
	inflationary models that naturally lead to features in the scalar
	power spectrum.
	
	Inflation driven by two or more scalar fields offer a richer dynamics and phenomenology (see~\cite{Wands:2007bd} for a review).
With more than one field, there are several possibilities to produce features in the PPS, as first pointed out in ~\cite{Kofman:1988xg}. Beyond introducing features in the potential for the inflatons~\cite{Price:2014xpa}, other 
phenomena are at play, such as particle production of additional spectator (i.e. with zero vev) fields coupled to the inflaton~\cite{Chung:1999ve}, couplings to gravity \cite{Pi:2017gih,Mori:2017caa,Gundhi:2018wyz} or a turn in the trajectory in field space.
Features originating from a turn in the trajectory in field space have been studied explicitly including curvature and isocurvature perturbations~\cite{Gao:2012uq} or 
in an effective single field approach by integrating out heavy degrees of 
freedom~\cite{Achucarro:2010da}.

In this paper we adopt the approach of a direct integration of two-field dynamics with turns in field space, trying to avoid complicated forms, such as allowing transitions, steps or cutoffs in the effective potential. Indeed, in two field models, deviations from slow-roll can be achieved between two different stages of inflation, as in double inflation~\cite{Polarski:1992dq}, and this is the case we will focus on. The possible connection between two field inflation and features at the largest scales has been already made in~\cite{Feng:2003zua}. 
	However, differently from~\cite{Feng:2003zua}, we demonstrate that a step-like feature with a higher amplitude on large scales is obtained when isocurvature perturbations are correctly taken into account,
which is not so interesting from an observational point of view for CMB temperature anisotropies. 
	
	We  therefore study the generation of features in a two 
	field model consisting of a canonical scalar field (say, $\phi$) and a second scalar field (say, $\chi$) with a non-canonical kinetic term of the form $f(\phi) (\partial\chi)^2$. For works on two field inflation with these non-canonical kinetic terms - not necessarily connected to features - see for example \cite{Berkin:1991nm,Starobinsky:1994mh,GarciaBellido:1995fz,Starobinsky:2001xq,DiMarco:2002eb,DiMarco:2005nq,Lalak:2007vi}. 
	Thereby, we focus on a setting in which $\phi$ ($\chi$) is the effectively heavier (lighter) field driving the first (second) stage of inflation. 
	 Since the deviations from slow-roll do not permit in general analytical calculations, we therefore use complete numerical computations to determine the primordial power spectra by evolving both background and perturbations.  
	 We  show that the effective mass of isocurvature perturbations grows with 
	 the coupling $f(\phi)$ during the first stage of inflation. Therefore, when isocurvature perturbations cross the Hubble radius, they decay and the feedback to curvature perturbation is thus suppressed. This effect reduces the correlation between the curvature and the 
	isocurvature perturbation in a way that a suppression of power is 
	achieved for scales that cross the Hubble radius before the temporary violation of the slow-roll conditions, that is during the first stage of inflation.

	This paper is organized as follows. 
	In the following section, we  describe the model of our interest
	and summarize the equations governing the background and the perturbations. 
	In Sec.~\ref{sec:formalism}, we  outline the numerical procedure 
	that we  adopt to evolve the background and the perturbations. In particular, we  describe in Sec.~\ref{sec:Bo-Feng} the important role isocurvature perturbations play in
our analysis. In  Sec.~\ref{sec:results}, we then introduce our toy model and present the results of our numerical computation for the PPS for a large range of scales and comment on the role of the different parameters at play in generating relevant features. We then discuss the two cases when scales relevant to CMB anisotropies observations cross the Hubble radius during or well before the slow-roll violation in Sec.~\ref{sec:CMB}.  In Appendix~\ref{appendix:VaryingMass}, for completeness, we provide an additional scan of the parameter space for the model presented in Sec.~\ref{sec:CMB}.
	
\section{Theoretical construction}~\label{sec:formalism}
	
In this section, we review the formalism essential to study
the generation of perturbations in two field models of inflation with a coupling between the kinetic terms in terms of tangent and orthogonal field ~\cite{Gordon:2000hv,GrootNibbelink:2001qt,DiMarco:2002eb,DiMarco:2005nq}.

\subsection{Background Dynamics}
	
	We  consider a model consisting of two scalar fields, say, 
	$\phi$ and $\chi$, whose dynamics is governed by the following 
	action:
	\begin{equation}
	\label{action}
	S[\phi,\chi]=\int \d^4x\,\sqrt{-g}\,
	\left[\frac{\Mpl^2}{2} R
	-\frac{1}{2}(\partial \phi)^2
	-\frac{f(\phi)}{2}(\partial\chi)^2-V(\phi,\chi)\right].
	\end{equation}
	Note that, while $\phi$ is a canonical scalar field, $\chi$ is 
	a non-canonical scalar field due to the presence of the function
	$f(\phi)$ in the term describing its kinetic energy.
	Evidently, apart from the potential $V(\phi,\chi)$, through which 
	the fields can in principle interact, the function $f(\phi)$ also leads to an
	interaction between the fields. In order to connect with the equations of Refs.~\cite{DiMarco:2002eb,DiMarco:2005nq} we define $f(\phi)\equiv{\rm e}^{2 b(\phi)}$.
	
	We  work with the spatially flat 
	Friedmann-Lema{\^i}tre-Robertson-Walker 
	(FLRW) universe described by the line-element 
	\begin{equation}
	\d s^2=-\d t^2+a^2(t)\,\d{\bm x}^2,\label{flrw} 
	\end{equation}
	where $a(t)$ is the scale factor and $t$ is the cosmic time.
	In such a smooth background, the equations of motion governing 
	the homogeneous scalar fields are given by
	\begin{subequations}
		\label{eq:sf}
		\begin{eqnarray}
		\label{eq:KGphi}
		\ddot{\phi}+3 H\dot{\phi}+V_\phi
		&=&b_\phi e^{2 b}\dot{\chi}^2,\\
		\label{eq:KGchi}
		\ddot{\chi}+(3 H+ 2 b_\phi \dot{\phi}) \dot{\chi}
		+{\rm e}^{-2 b}V_\chi&=&0,
		\end{eqnarray}
	\end{subequations}
	where, as usual, the overdots represent differentiation with respect
	to the cosmic time, $H=\dot{a}/a$ is the Hubble parameter, while the 
	subscripts denote differentation of the potential $V(\phi,\chi)$ and 
	the function $b(\phi)$ with respect to the corresponding fields.
	The dynamics of the scale factor is described by the following
	Friedmann equations:
	\begin{subequations}
		\label{eq:f}
		\begin{eqnarray}
		H^2&=&\frac{1}{3 \Mpl^2}		\left[\frac{\dot{\phi}^2}{2}		+{\rm e}^{2 b}\frac{\dot{\chi}^2}{2}+V\right],\\
		\dot{H}&=&-\frac{1}{2 \Mpl^2}\left[\dot{\phi}^2
		+{\rm e}^{2b}\dot{\chi}^2\right].
		\end{eqnarray}
	\end{subequations}

	To characterize the evolution of the background, it is useful 
	to introduce the so-called Hubble Flow Functions (HFFs) as 
	follows~\cite{Schwarz:2001vv}:
	\begin{equation}
	\epsilon_{i+1}\equiv\frac{\d \ln \epsilon_i}{\d N},
	\end{equation}	with
	\begin{equation}
    \epsilon_0\equiv\frac{H_{\textup{\rm in}}}{H}
    \end{equation}
	where $H_{\rm in}$ is the value of the Hubble parameter at some
	initial time during inflation, and $N=\int \d t H$ represents the number of e-folds. 
	The HFFs hierarchy is not only useful to define a successful  slow-roll regime (i.e. $\epsilon_1 \,, \epsilon_2 \ll 1$) or the end of inflation ($\epsilon_1=1$), but also to quantify violations of the slow-roll regime which are of particular interest for this paper.

\subsection{Linear Perturbations}	
	Let us now turn to the dynamics of perturbations around the homogeneous \emph{vev} of the scalar fields. 	To describe the scalar perturbations, for simplicity, we  work in the Newtonian gauge.
	Since scalar fields do not have anisotropic stress to linear order, in the Newtonian gauge, the perturbed FRLW metric takes the form \cite{Mukhanov:1990me}
	\begin{equation}
	\d s^2=-(1+2\Phi)\,\d t^2+a^2(t)\, (1-2\Phi)\,\d{\bm x}^2 \,,
	\end{equation}
	where $\Phi$ is the Bardeen potential characterizing the perturbations.
	
	To study the evolution of the perturbations, it proves to be convenient to decompose the perturbations in the scalar fields, say, $\delta\phi$
	and $\delta\chi$, along directions that are parallel and orthogonal 
	directions to the trajectory in the field space~\cite{Gordon:2000hv}.
	These correspond to the instantaneous adiabatic and isocurvature 
	perturbations and they are given by the expressions~\cite{DiMarco:2002eb}
	\begin{subequations}
		\label{eq:ai}
		\begin{eqnarray}
		\label{adi}
		\delta\sigma&=&\cos\theta\, \delta\phi
		+\sin\theta\, {\rm e}^b\, \delta\chi,\\
		\label{iso}
		\delta s&=&-\sin\theta\,\delta\phi
		+\cos\theta\, {\rm e}^b\, \delta\chi,
		\end{eqnarray}
	\end{subequations}
	with the quantity $\theta$ being an angle in the field space that is 
	defined through the relations
	\begin{equation}
	\cos\theta=\frac{\dot{\phi}}{\dot{\sigma}},\quad
	\sin\theta={\rm e}^b \frac{\dot{\chi}}{\dot{\sigma}},\quad
	\dot{\sigma}^2=\dot{\phi}^2+{\rm e}^{2b}\dot{\chi}^2.
	\end{equation}
	The equations~(\ref{eq:sf}) describing background scalar fields
	can be combined to arrive at the following equations governing 
	$\dot{\sigma}$ and the angle $\theta$:
	\begin{subequations}
		\begin{eqnarray}
		\label{KGsigma}
		\ddot{\sigma}+3 H \dot{\sigma}+V_\sigma&=&0,\\
		\dot{\theta}&=&-\frac{V_s}{\dot{\sigma}}-b_\phi\dot{\sigma}\sin\theta
		\end{eqnarray}
	\end{subequations}
	where $V_\sigma$ and $V_s$ are given by 
	\begin{subequations}
		\begin{eqnarray}
		V_\sigma&=&V_\phi\cos\theta+{\rm e}^{-b}\,V_\chi\sin\theta,\\
		V_s&=&-V_\phi\sin\theta+{\rm e}^{-b}\,V_\chi\cos\theta.
		\end{eqnarray}
	\end{subequations}
	 We also introduce here the following quantities for future convenience
	\begin{subequations}
		\begin{eqnarray}
		V_{\sigma\sigma} &=&V_{\phi\phi}\cos^2\theta 
		+ {\rm e}^{-b} V_{\phi\chi} \sin 2 \theta
		+ {\rm e}^{-2 b} V_{\chi\chi} \sin^2\theta,\\
		V_{ss} &=&V_{\phi\phi}\sin^2\theta
		- {\rm e}^{-b} V_{\phi\chi}\sin 2 \theta
		+ {\rm e}^{-2 b} V_{\chi\chi}\cos^2\theta ,\\
		V_{\sigma s} &=&-V_{\phi\phi}\cos\theta\sin\theta
		+ {\rm e}^{-b} V_{\phi\chi} (\cos^2\theta-\sin^2\theta)
		+ {\rm e}^{-2 b} V_{\chi\chi}\cos\theta\sin\theta.
		\end{eqnarray}
	\end{subequations}

	To characterize the perturbations, we consider the gauge-invariant Mukhanov-Sasaki variable $Q_\sigma$ associated to $\delta \sigma$, i.e. $Q_\sigma = \delta \sigma + \dot \sigma/H \Phi$, and $\delta s$ (which is already gauge-invariant), associated with the adiabatic and isocurvature fields.
	The equations of motion governing $Q_\sigma$ and $\delta s$ can be arrived 
	at from the differential equations describing the perturbations in the 
	scalar fields and the first order Einstein's equations.
	They can be obtained to be 
	\begin{subequations}
		\label{eq:Q}
		\begin{eqnarray}
		\ddot{Q}_\sigma + 3 H \dot{Q}_\sigma
		&+&\l[\f{k^2}{a^2}+V_{\sigma\sigma}-\dot{\theta}^2
		-\f{1}{a^3 \Mpl^2}
		\l(\f{a^3\dot{\sigma}^2}{H}\r)^{\fatdot}
		+b_\phi u(t)\r]Q_\sigma\nn\\
		&=& 2\l(\dot{\theta} \delta s \r)^{\fatdot}
		-2\left(\frac{\dot{H}}{H}
		+\frac{V_\sigma}{\dot{\sigma}}\right)\dot{\theta} \delta s
		+  b_{\phi\phi}\dot{\sigma}^2\sin2\theta \delta s +2 b_\phi h(t)  \,,\qquad\\
		\ddot{\delta s} + 3 H\dot{\delta s}
		&+&\l[\f{k^2}{a^2}+V_{ss}+3\dot{\theta}^2
		+b^2_\phi g(t)+b_\phi f(t)
		-b_{\phi\phi}\dot{\sigma}^2
		-4\f{V_s^2}{\dot{\sigma}^2}\r] \delta s \nn\\
		&=& 2\f{V_s}{H}\l(\f{H}{\dot{\sigma}}Q_\sigma\r)^{\fatdot},
		\end{eqnarray}
	\end{subequations}
	where the quantities $u(t)$, $h(t)$, $g(t)$ and $f(t)$ are given by
	\begin{subequations}
		\begin{eqnarray}
		u(t)&=&\dot{\theta}\dot{\sigma}\sin\theta-{\rm e}^{-b}V_\chi\sin\theta\cos\theta,\\
		h(t)
		&=&-\dot{\sigma} (\sin\theta\, \delta s)^{\fatdot}
		-\sin\theta \left(\frac{\dot{H}}{H}\dot{\sigma}
		+2 V_\sigma\right) \delta s-3 H\dot{\sigma}\sin\theta \delta s,\\
		g(t)&=&-\dot{\sigma}^2(1+3\sin^2\theta),\\
		f(t)&=&V_\phi(1+\sin^2\theta)-4 V_s\sin\theta.
		\end{eqnarray}
	\end{subequations}

	We choose to impose the following Bunch-Davies 
	initial conditions on the variables $Q_\sigma$ and $\delta s$ 
	at early times when the modes of cosmological interest are well 
	inside the Hubble radius:
	\begin{equation}
	\label{eq:ic}
	Q_\sigma(\tau)
	\simeq \delta s(\tau)
	\simeq \f{1}{a(\tau)}\,\f{{\rm e}^{\imath k \tau}}{\sqrt{2 k}},
	\end{equation}
	where $\tau$ is the conformal time coordinate defined as $\tau=
	\int\, \d t/a(t)$. We have verified that the same results are obtained by considering Bunch-Davies 
	initial conditions for the gauge-invariant variables associated to $\phi$ and $\chi$ instead. Note, however, that extra care is needed when imposing the initial conditions in case the turning rate, i.e. $\dot{\theta}$, is not small  and $Q_\sigma$ and $\delta_s$ can be correlated at early times \cite{Cremonini:2010ua}.
	
	While $Q_\sigma$ and $\delta s$ are convenient variables to identify the most natural initial conditions that can be imposed, it is also useful to think in terms of curvature and 
	isocurvature fluctuations, which are used to classify initial conditions in the post-inflationary expansion.
		These quantities are given in terms of the variables $Q_\sigma$ and $\delta s$ by the following relations:
	\begin{equation}
	\mathcal{R}= \frac{H}{\dot{\sigma}}Q_\sigma,\quad
	\mathcal{S}=\frac{H}{\dot{\sigma}} \delta s.
	\end{equation}

	Upon utilizing the equations~\eqref{eq:Q}, it is straightforward
	to arrive at the following equations governing the curvature
	and the isocurvature perturbations $\cR$ and $\cS$:
	\begin{subequations}
		\label{eq:RS}
		\begin{eqnarray}
		\label{eq:R}
		\ddot{\mathcal{R}}
		&+&\left(H+2\f{\dot{z}}{z}\r)\dot{\mathcal{R}}
		+\frac{k^2 }{a^2}\mathcal{R}
		=-\frac{2 V_s}{\dot{\sigma}}\dot{\mathcal{S}}
		-\,2\biggl(-{\rm e}^{-b}  b_{\phi }\cos^2\theta\, V_{\chi }
		+ \sin\theta b_{\phi } V_{\sigma}\nn\\
		& &\qquad\qquad\qquad\qquad\qquad\;\;
		+\, V_{\sigma s}+\f{\dot{\sigma}}{H M^2_\textup{pl}}V_s\biggr)\mathcal{S},\\
		\ddot{\mathcal{S}}
		&+&\l(H+2\f{\dot{z}}{z}\right)\dot{\mathcal{S}}
		+\biggl\{\frac{k^2}{a^2}-2 H^2-\dot{H}+\frac{H \dot{z}}{z}
		+\frac{\ddot{z}}{z}-\dot{\theta}^2-\dot{\sigma}^2 b_{\phi }^2 \cos^2\theta
		-\dot{\sigma}^2 b_{\varphi \varphi }+V_{ss}\nn\\
		&+&\,b_{\phi } \l[4 \sin \theta \, V_s
		+(1+\sin^2\theta) V_{\phi }\r]
		\biggr\}\cS
		=\f{2  V_s}{\dot{\sigma}}\dot{\mathcal{R}},
		\end{eqnarray}
	\end{subequations}
	where $z\equiv a \dot{\sigma}/H$.

Once the potential $V(\phi,\chi)$, the function $b(\phi)$, and the 
values of the parameters describing them are specified, we  
first integrate the equations~(\ref{eq:sf}) and~(\ref{eq:f}) to 
arrive at the background quantities. 
As is often done in the case of inflation, in order to efficiently
integrate the equations involved, we work with the number of 
e-folds~$N$ as the independent time variable. 
In all the different models that we  consider, we  assume 
that the pivot scale $k=0.05 \,\text{Mpc}^{-1}$ leaves the Hubble 
radius at $50$ e-folds before the end of inflation. 

With the background quantities in hand, we  then go on to integrate 
the equations~(\ref{eq:RS}) governing the curvature and isocurvature perturbations~$\cR$ and~$\cS$.
Following Refs.~\cite{Tsujikawa:2002qx,Lalak:2007vi,vandeBruck:2014ata}, we  integrate the equations~(\ref{eq:RS}) by imposing the Bunch-Davies initial condition~(\ref{eq:ic}) on $Q_\sigma$ and 
assuming the initial value of $\delta s$ to be zero.
We  then integrate the equations a second time by interchanging the initial conditions on~$Q_\sigma$ and~$\delta s$. This ensures no correlations between the curvature and isocurvature fluctuations when the modes are deep inside the Hubble radius. 
If we denote these two sets of solutions as ($\cR_{1}$, $\cS_1$) 
and ($\cR_2$, $\cS_2$), then the power spectra describing the curvature and the isocurvature perturbations as well as the cross-correlations between them are defined respectively as
\begin{subequations}
	\label{P1P2}
	\begin{eqnarray}
	\mathcal{P}_{\cR}(k)
	&=&\frac{k^3}{2\pi^2}
	\l(\lvert\cR_1\rvert^2+\lvert\cR_2\rvert^2\r)
	=\mathcal{P}_{\cR_1}(k)+\mathcal{P}_{\cR_2}(k),
	\label{eq:PR}\\
	\mathcal{P}_{\cS}(k)
	&=&\frac{k^3}{2\pi^2}
	\l(\lvert\cS_1\rvert^2+\lvert\cS_2\rvert^2\r),\\
	\mathcal{C}_{\cR\cS}(k)
	&=&\frac{k^3}{2\pi^2}
	\l(\cR^\ast_1\cS_1+\cR^\ast_2\cS_2\r).
	\end{eqnarray}
\end{subequations}

		\section{The importance of isocurvature perturbations}\label{sec:Bo-Feng}	

\begin{figure}
 	\begin{center} 
 		\resizebox{218pt}{150pt}{\includegraphics{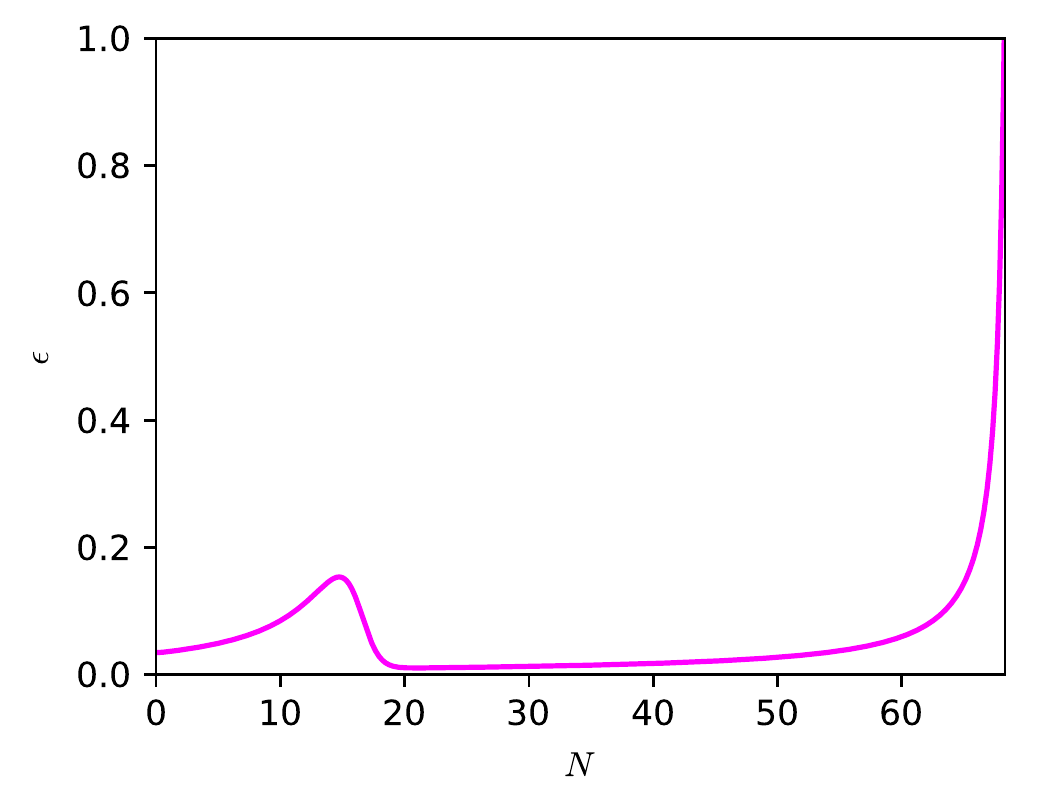}}
 		\resizebox{218pt}{150pt}{\includegraphics{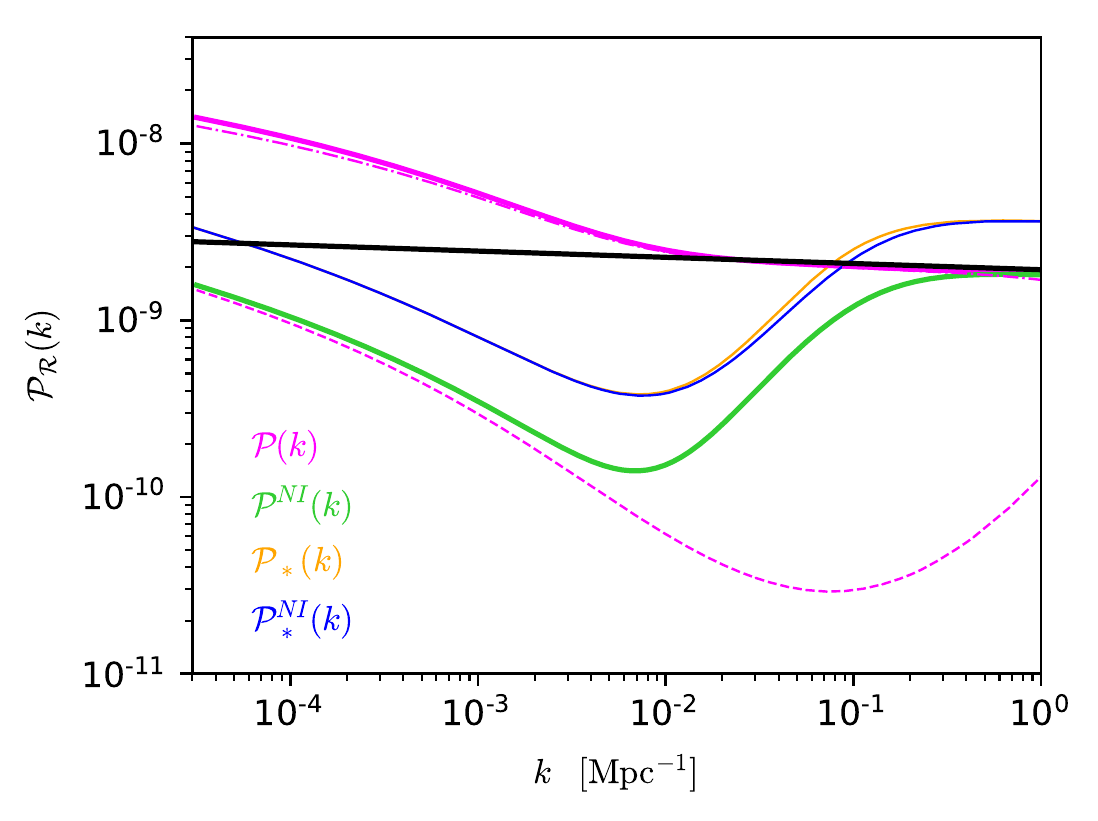}}
 	\end{center}
 	\caption{\label{fig:BoFeng} Left: Evolution of the slow-roll parameter $\epsilon$ for the double inflation model in Eq.~\eqref{eqn:DoubleInflation} with $\chi_0=6m_\phi$. Right: The green line represents the scalar power spectrum found in Ref.~\cite{Feng:2003zua} where the effect of isocurvature perturbation was neglected.		
 	See the magenta solid line for the correct result. 
 	In more detail, the magenta dashed (dotted) line shows the scalar power spectrum for $\mathcal{R}_1$ ($\mathcal{R}_2$). 
 	We also plot the power spectra at Hubble crossing computed taking (not taking) into account isocurvature perturbations in orange (blue) lines.  		}    
 \end{figure}

	In this section, we  provide an explicit example of how it is difficult to achieve a suppression of the spectrum of curvature perturbations at the largest scales when inflation is driven by two canonical massive scalar fields. The interest in this double inflation model ~\cite{Silk:1986vc,Polarski:1992dq,Polarski:1994rz,Lesgourgues:1997ki}
	 \begin{equation}
 \label{eqn:DoubleInflation}
 V(\phi,\,\chi)=\frac{m^2_\phi}{2}\phi^2+\frac{m^2_\chi}{2}\chi^2,
 \end{equation}
 in the context of features has been driven by Ref.~\cite{Feng:2003zua}.  In particular, it was claimed that this model can lead to a suppression of power in the primordial power spectrum~\cite{Feng:2003zua}. The deviation from the nearly scale-invariant result is achieved due to a temporary violation of slow-roll (in the sense discussed in section~\ref{sec:formalism}) at the transition between the first period of inflation driven by the heavier field  ($m_\phi=6\,\chi_0$) and the second one driven by the lighter field~$\chi$, which results in a local bump in the first slow-roll parameter $\epsilon$, as can be seen in the left panel of Fig.~\ref{fig:BoFeng}.

  The curvature power spectrum obtained in~\cite{Feng:2003zua} is plotted in the right panel of Fig.~\ref{fig:BoFeng} in green line. However this result is obtained by neglecting the coupling between $\mathcal{R}$ and $\mathcal{S}$ in Eq.~\eqref{eq:R}. Taking into account the full evolution equations for $\mathcal{R}$ leads instead to the magenta solid power spectrum in the left panel of  Fig.~\ref{fig:BoFeng}, i.e. to a nearly scale-invariant power spectrum on small scales with a bump on large scales. We also plot the power spectra at Hubble crossing computed by taking (not taking) into account isocurvature perturbations in orange (blue) lines: as expected, they are nearly the same, as the isocurvature sourcing inside the Hubble radius is negligible. 
  We also plot in the right panel of  Fig.~\ref{fig:BoFeng} the curvature spectra corresponding to the two $\mathcal{R}_1$ and $\mathcal{R}_2$, computed properly taking into account the isocurvature perturbations. As it is easy to see, in this case, the full power spectrum is almost given by $\mathcal{P}_{\mathcal{R}_1}$ only, as we will show also for  the model in the next Section. 
 
 The importance of isocurvature perturbations in the context of features in the PPS generated by a sharp turn in field space is not only limited  to the model in Eq. (3.1) but is more general. See for instance Ref. \cite{Lalak:2007vi,vandeBruck:2014ata} for the inclusion of isocurvature perturbations in Roulette Inflation \cite{Bond:2006nc}.

	\section{Generating features in the primordial power spectrum by a coupling between kinetic terms }\label{sec:results}

	In this section, we introduce a model in which primordial features at large scales of the desired form are more easily produced. 
		We  assume that the non-canonical function $b(\phi)$ has the form $b(\phi)=b_1 \phi$, where $b_1$ is a constant with dimensions $M^{-1}$. 
	We will see that the above simple form for the function $b(\phi)$ can lead to several type of features which could be observationally relevant for CMB anisotropies.  Note that with this choice, the field space has an hyperbolic structure with a constant and negative Ricci curvature $R=-2 db/d\phi^2=-2 b_1^2$.

	Although the mechanism we present is quite general for those 
	initial conditions in which inflation is first driven by a field $\phi$ with an effective mass which is larger than $\chi$, we consider the following potential:
	\begin{equation}
	\label{eq:potential}
		V(\phi,\,\chi)=\frac{m^2_\phi}{2}\phi^2+V_0\frac{\chi^2}{\chi_0^2+\chi^2} \,.
		\end{equation}
We consider a KKLTI-like potential \cite{Kallosh:2018zsi} 
instead of a simplest mass term for $\chi$ in order to have a tensor-to-scalar ratio compatible with the most recent
constraints \cite{Akrami:2018odb,Aghanim:2018eyx}.

\begin{figure}
		\begin{center} 
			\resizebox{214pt}{172pt}{\includegraphics{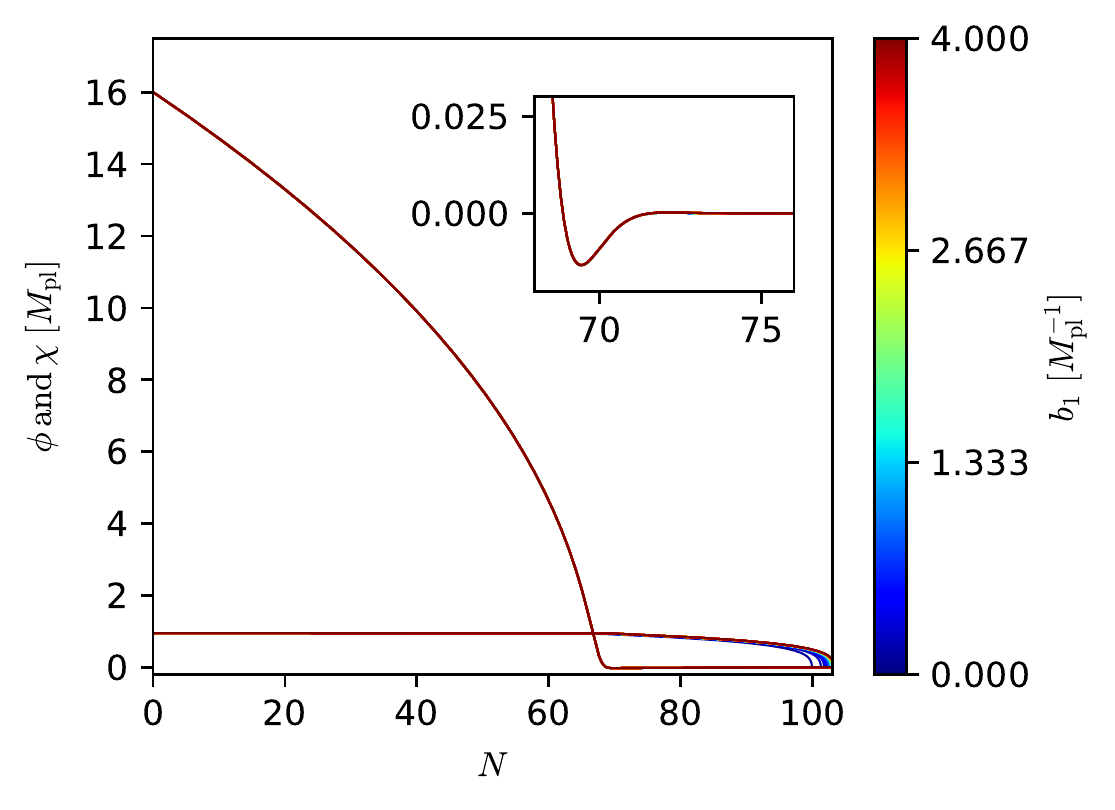}}			\resizebox{214pt}{172pt}{\includegraphics{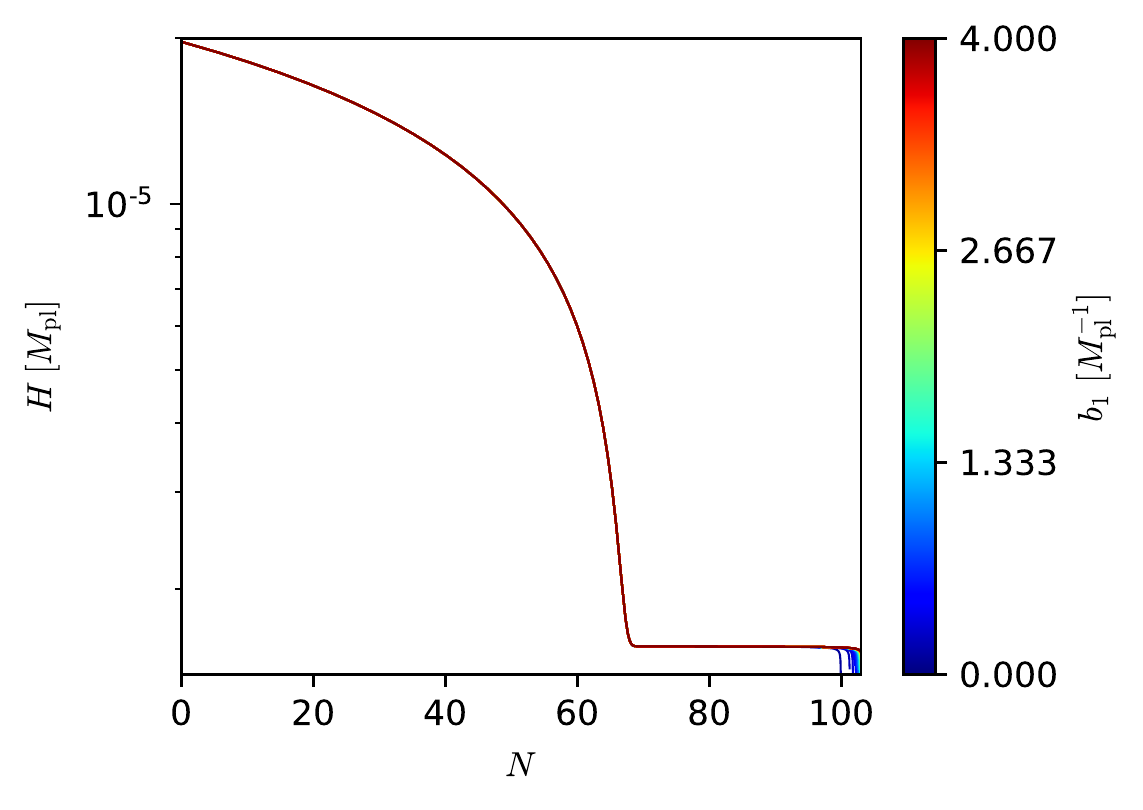}}	
	
	\resizebox{214pt}{172pt}{\includegraphics{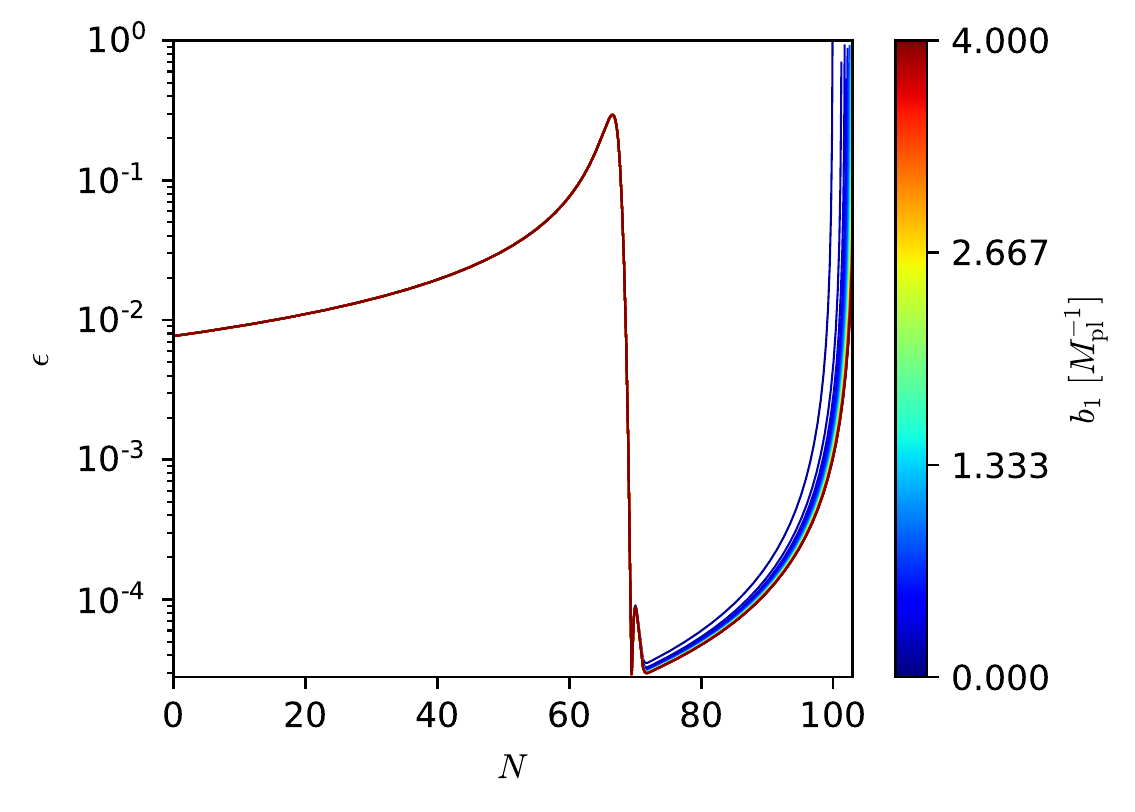}}
			\resizebox{214pt}{172pt}{\includegraphics{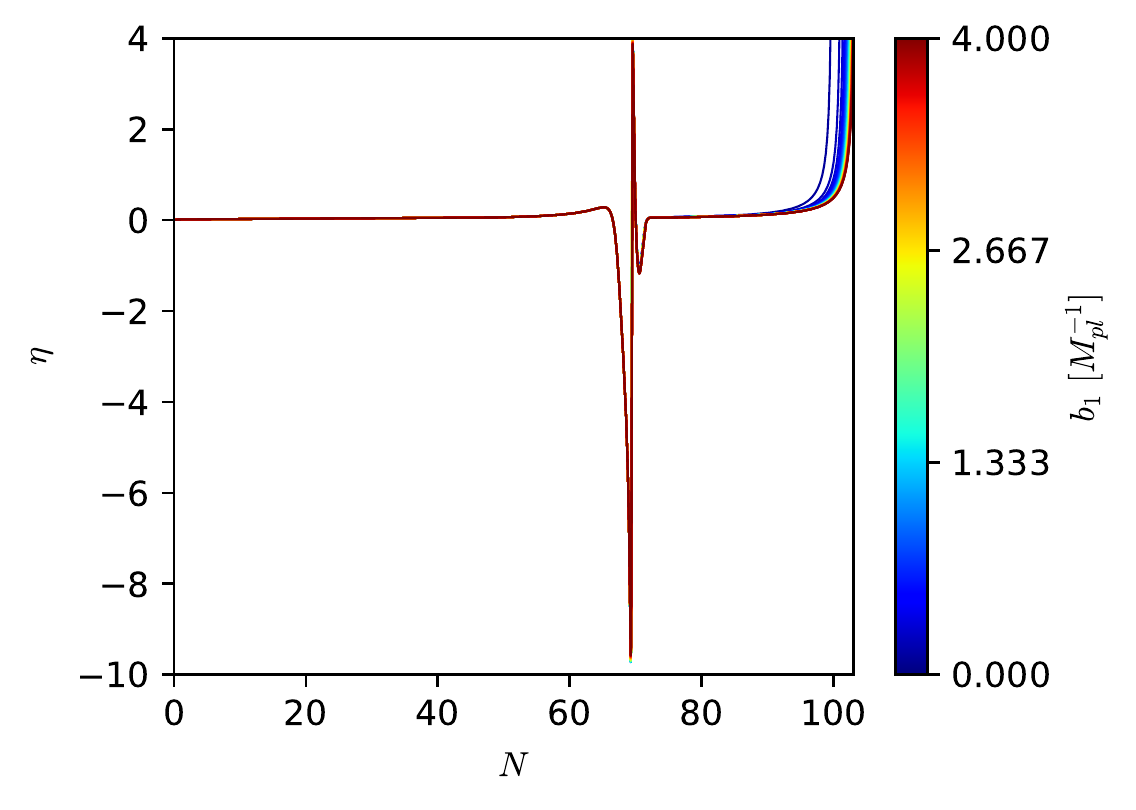}}	
		\end{center}
		\caption{\label{fig:Background}[Top] Evolution of the scalar fields (left) and the Hubble parameter $H$ (right). [Bottom] Evolution of the first two slow-roll parameters $\epsilon$ (left) and $\epsilon_2\equiv\eta$ (right).  The ratio $(m_\phi\,M_\textup{pl})^2 /V_0=1.2$ is fixed and  $b_1$ is varied for a continuous range of values.} 
	\end{figure}

	In Fig.~\ref{fig:Background}, we plot relevant background quantities 
	for the illustrative case of $(m_\phi\,M_\textup{pl})^2 /V_0=1.2$, $\chi_0=0.06 M_\textup{pl}$. We vary $b_1$ in order to highlight the effects of the coupling with the $\chi$ kinetic term. 
We choose $\phi_i=16 \,M_\textup{pl}$ and $\chi_i=0.96\,M_\textup{pl}$ for the initial values of the scalar fields and we fix their initial time derivatives by imposing slow-roll initial conditions on $\dot{\phi}_i$ and $\dot{\chi}_i$. Note that this is not strictly necessary as the fields relax to the slow-roll attractors even if the velocities are chosen randomly. We  consider sets of parameters that are relevant for observations in the next section.    
	 
	 As can be easily seen from Fig.~\ref{fig:Background}, the heavier of the two fields, i.e. $\phi$, rolls down its potential driving a first phase of inflation  while the lighter field $\chi$ remains frozen (see Ref.~\cite{Braglia:2020eai} for a detailed analytical description of the background evolution). This dynamics, also typical of multifield $\alpha$-attractor models \cite{Linde:2018hmx,Iarygina:2018kee}, can be easily understood as follows. For $b_1>0$ and the initial values of $\phi$ considered in Fig.~\ref{fig:Background}, the term $\exp(2 b(\phi))V_\chi$ can be safely taken as zero\footnote{Note that, with our choice of $\chi_0=0.06 M_\textup{pl}$, we have $V(0,\,\chi)\simeq V_0$ during the first stage of inflation, and therefore $\chi\simeq\chi_i$ also for $b_1=0$. } as a good approximation and a solution of Eq.~\eqref{eq:KGchi} is $\chi_1=\chi_i$, where the subscript $1$ is to specify that this solution is valid during the first stage of slow-roll. Therefore, the term proportional to $\dot{\chi}^2$ in Eq.~\eqref{eq:KGphi}, along with $\ddot{\phi}$, can be discarded and the slow-roll solution to Eq.~\eqref{eq:KGphi} becomes simply: 
	 	\begin{equation}
	\phi_1(N)=\sqrt{\phi_i^2 +4 M_\textup{pl}^2 N}. 
	\end{equation}

	  The first stage of inflation dominated by $\phi$ eventually ends and, after a few damped oscillations, the pattern of which depends on the potential ratio $(m_\phi\,M_\textup{pl})^2 /V_0$, see also Appendix~\ref{appendix:VaryingMass},  it settles to the effective minimum of its potential which is found by solving 
	 \begin{equation}
	    V_\phi=b_\phi e^{2b}\dot{\chi}^2\simeq b_\phi \frac{e^{-2b}}{3}\left(\frac{V_\chi}{V(0,\,\chi)}\right)^2,
	 \end{equation}
	 which has a solution
	  \begin{equation}
	    \phi_\textup{min}\simeq\textup{const}\simeq \frac{\mathcal{W}\left(\frac{8 b_1^2 \chi_0^4}{3 \chi_i^6}\right)}{2 b_1}
	  \end{equation}
	  where $\mathcal{W}$ is the D'Alambert W function and we have assumed $\chi\sim\chi_i$. 
	 The field $\chi$ then starts a second inflationary phase. The Klein-Gordon equation \eqref{eq:KGchi} can be easily integrated to get   
	    \begin{equation}
    \chi_2(N)=\sqrt{-\chi_0^2+\sqrt{-8 (N-N_1)\chi_0^2+\left(\chi_i^2+\chi_0^2\right)^2 }}e^{-2 b_1\phi_\textup{min}},
    \end{equation}
    where we have denoted by $N_1$ the beginning of the second stage of inflation.
With this choice of parameters, although $\phi_\textup{min}$ is  small, the non-canonical kinetic term affects the background dynamics in making the second inflationary phase longer, as the third term in  Eq.~\eqref{eq:KGchi} acts as a friction term for the $\chi$ field exponentially, and to suppress $V_\chi$. As can be seen from the central panels of Fig.~\ref{fig:Background}, the first slow-roll parameter $\epsilon$ shows a bump between the two phases and the slow-roll conditions are violated, i.e. $\epsilon_2\equiv\eta>1$.

		\begin{figure}
		\begin{center} 
			\resizebox{143pt}{115pt}{\includegraphics{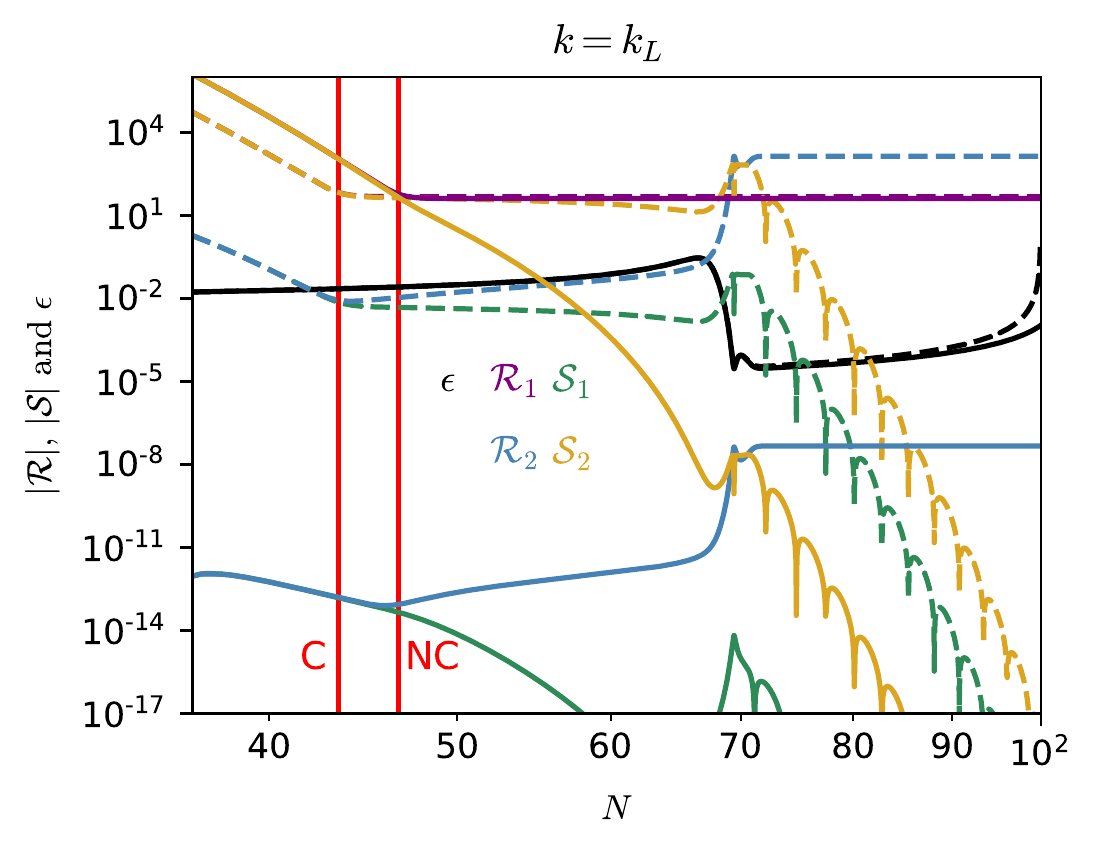}}		\resizebox{143pt}{115pt}{\includegraphics{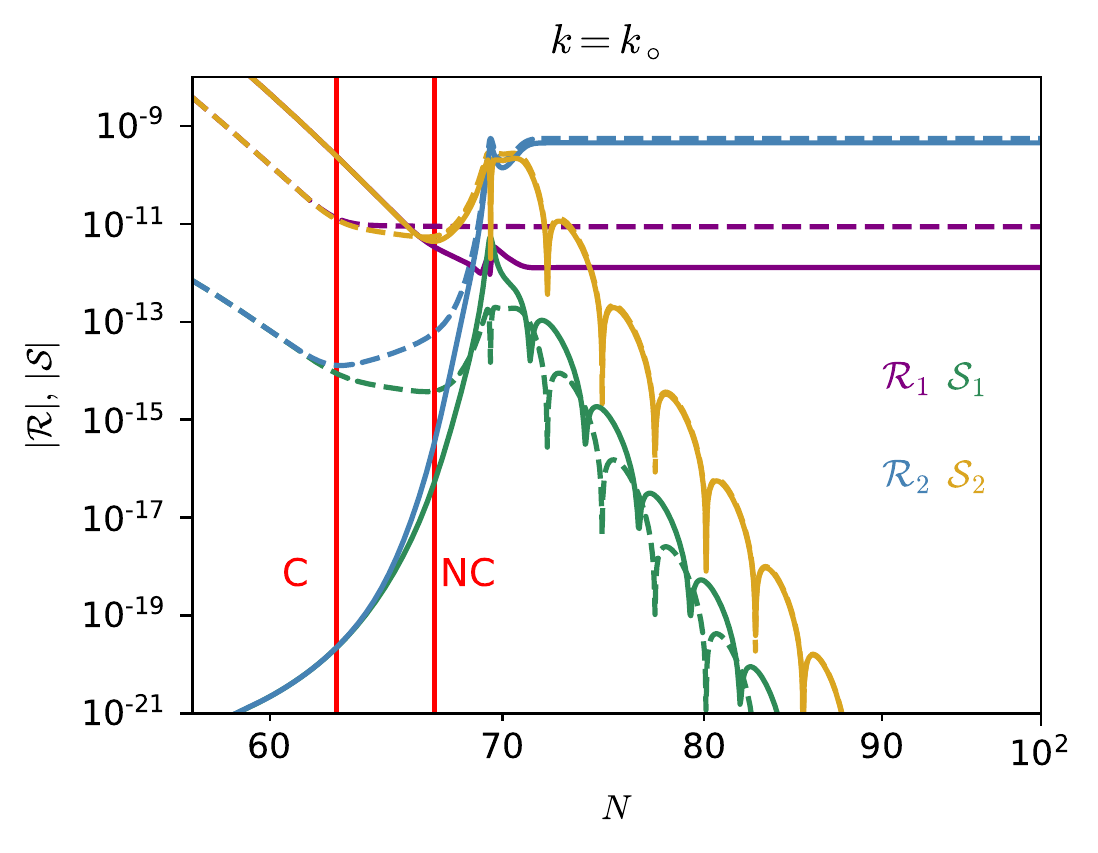}}		\resizebox{143pt}{115pt}{\includegraphics{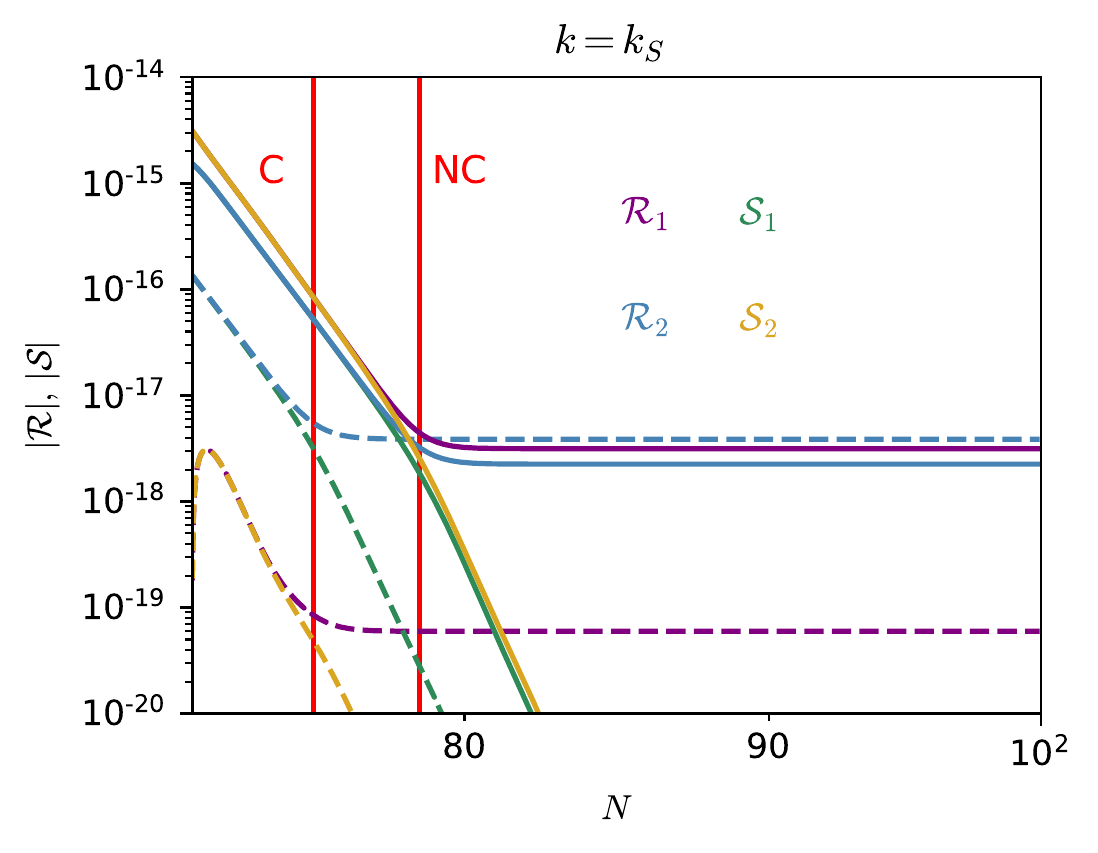}}	
		\end{center}
		\caption{\label{fig:modeEvolution} Evolution of the perturbed modes for a canonical coupling between the two scalar fields ($b_1=0$, dashed lines) and a non-canonical coupling with $b_1=4/M_{\mathrm pl}$ (continuous lines). We plot the evolution for the modes with $k_L/k_\circ \equiv 10^{-7}$ [Left], $k_\circ$ [Center] and $k_S/k_\circ \equiv 10^{5}$ [Right]. The red vertical lines signal the $e$-folds when the modes cross the Hubble radius. We also plot the $\epsilon$ parameter in black lines. The ratio $(m_\phi\,M_\textup{pl})^2 /V_0=1.2$ is fixed.}
	\end{figure}
	
	\begin{figure}
		\begin{center} \resizebox{443pt}{160pt}{\includegraphics{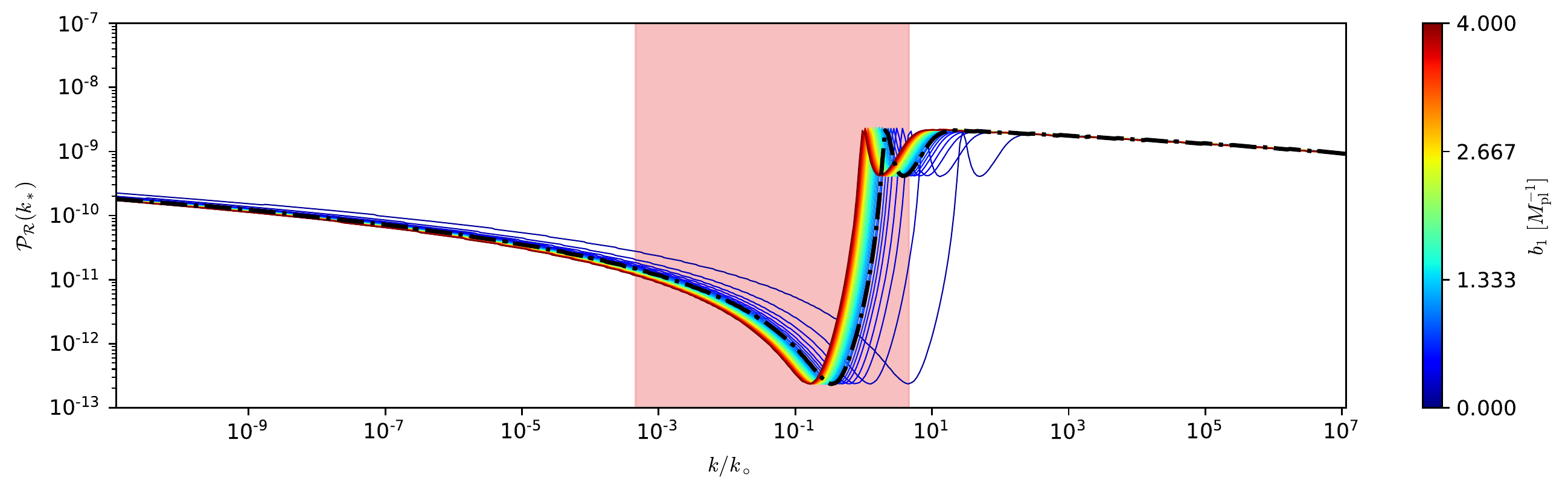}}
		\resizebox{443pt}{160pt}{\includegraphics{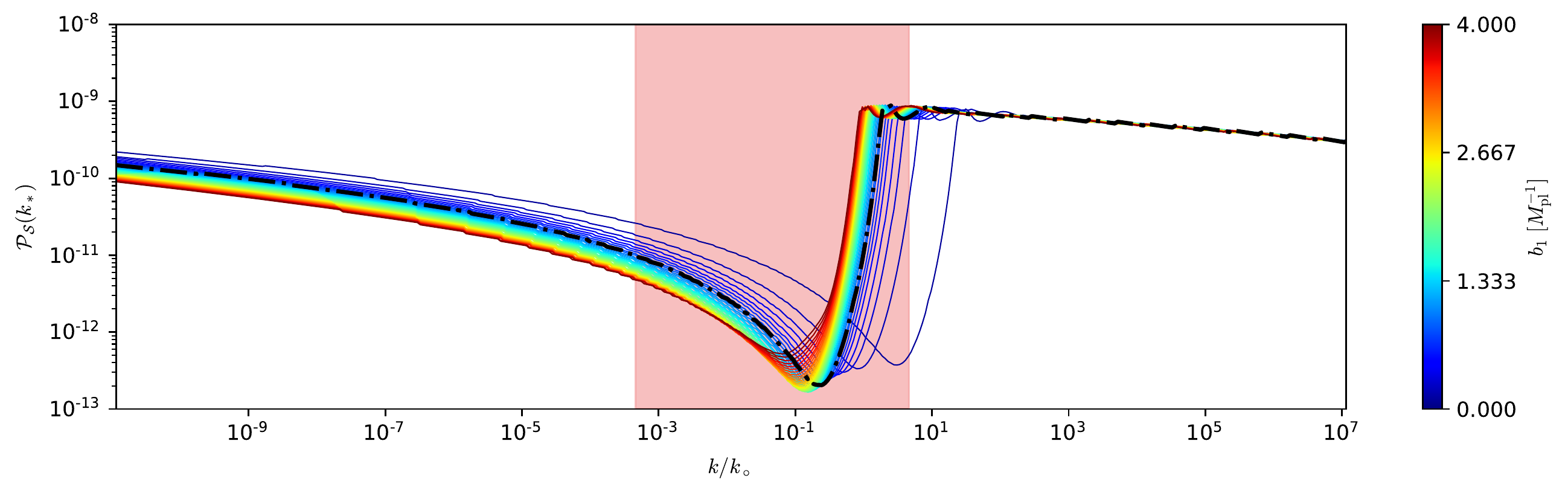}}
        \resizebox{443pt}{160pt}{\includegraphics{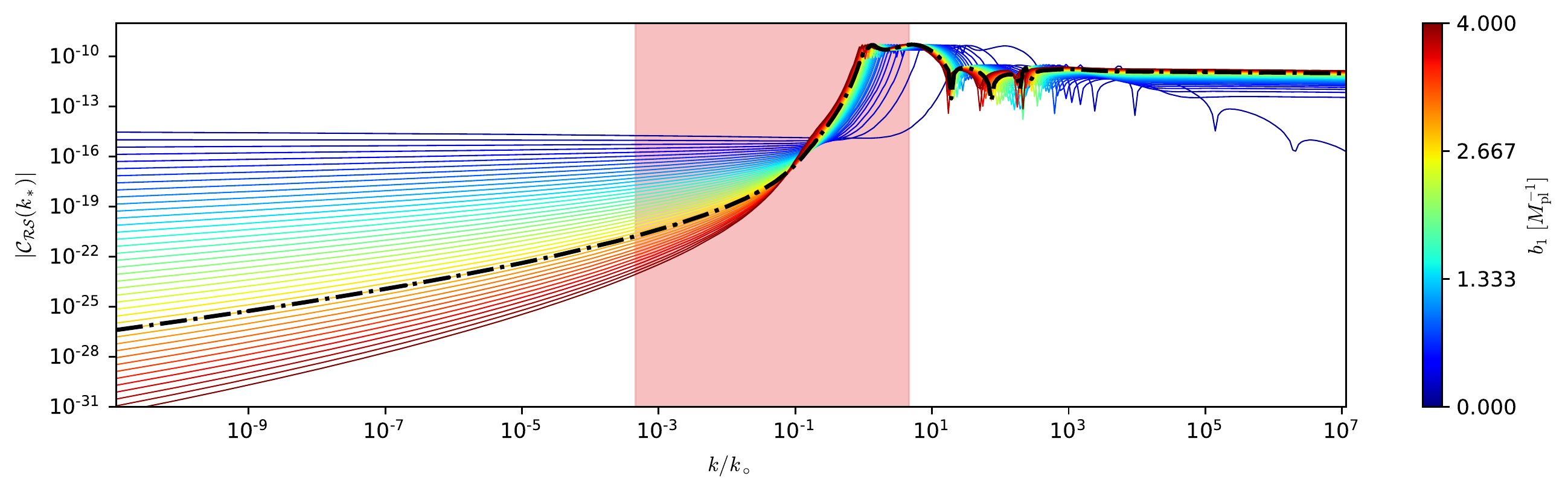}}
			\caption{\label{fig:power_Hubble}[Top] Curvature , [center] isocurvature, [bottom] curvature-isocurvature cross-correlation  power spectra at Hubble crossing. The ratio $(m_\phi\,M_\textup{pl})^2 /V_0=1.2$ is fixed and  $b_1$ is varied for a continuous range of values.  The $x$-axis is normalized to $k_\circ$, i.e. the scale which crosses the Hubble radius when $\epsilon$ reaches is maximum value in the dotted black model. The red band represent the range of scales that cross the Hubble radius during the range of $e$-folds when slow-roll is violated.} 		
		\end{center} 
	\end{figure}

		\begin{figure}
		\begin{center} 
            \resizebox{453pt}{170pt}{\includegraphics{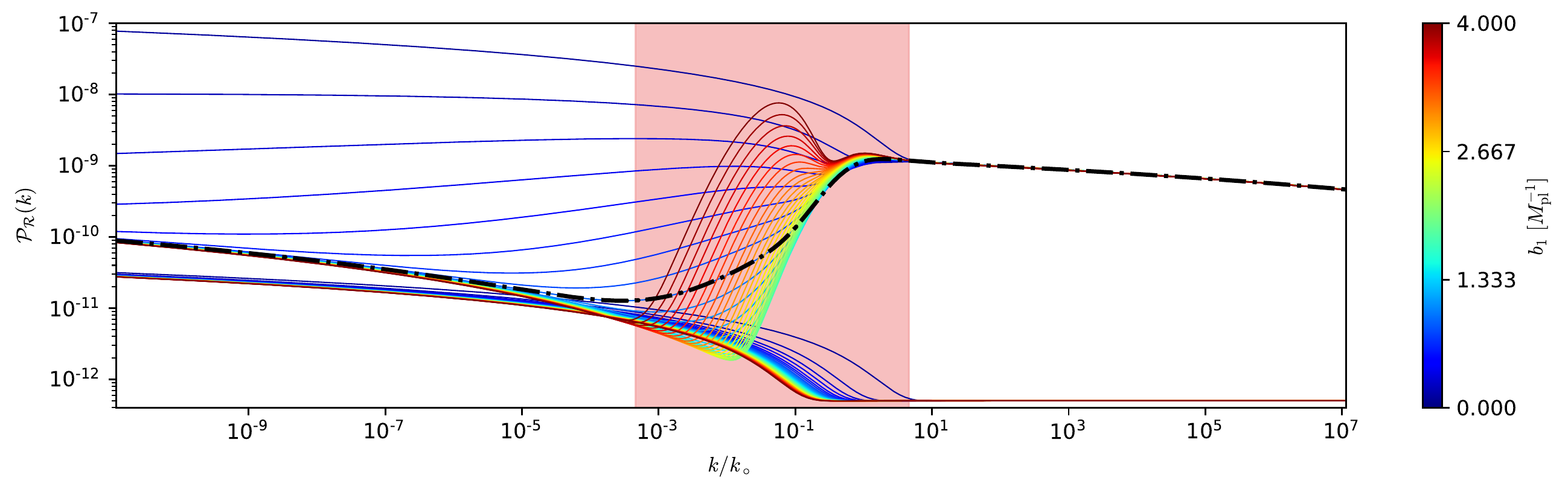}}
			\resizebox{453pt}{170pt}{\includegraphics{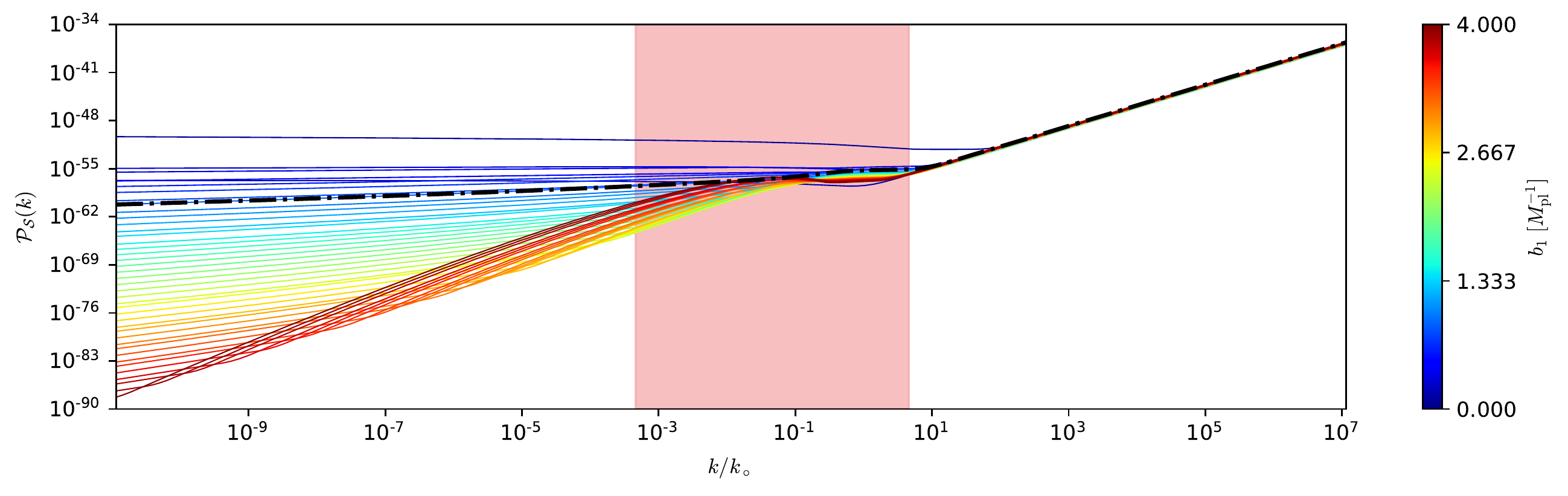}}
			\resizebox{453pt}{170pt}{\includegraphics{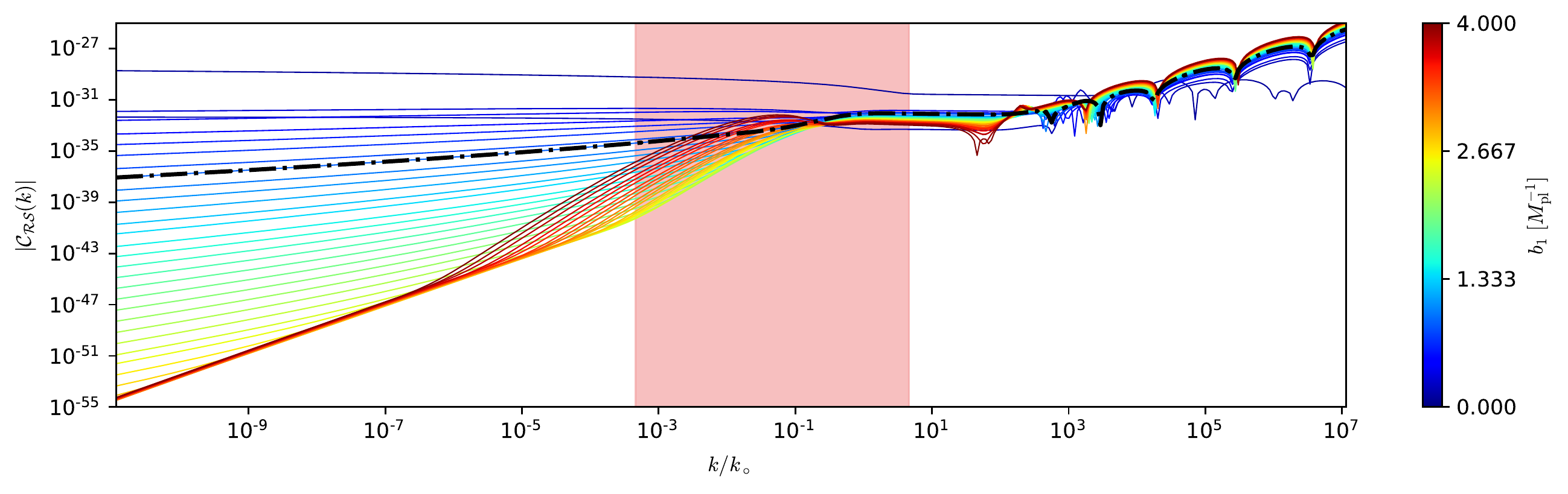}}
			\caption{\label{fig:power} [Top] Curvature , [center] isocurvature, [bottom] curvature-isocurvature cross-correlation  power spectra 
			evaluated at the end of inflation. The ratio $(m_\phi\,M_\textup{pl})^2 /V_0=1.2$ is fixed and  $b_1$ is varied over a continuous range of values.  In the top panel we also plot the tensor power spectrum evaluated at the end of inflation.} 	
		\end{center} 
	\end{figure}
	
	The evolution of fluctuations  is shown in Fig.~\ref{fig:power} for
three different modes that cross the Hubble radius well before ($k=k_L$, large scale mode), at ($k=k_\circ$, transition mode) and after ($k=k_S$, small scales modes) the transition between the 
	 	 two stages of inflation. Solid (dashed) lines represent  modes for 
	 	 $b_1=4 / M_\mathrm{pl}$ ($b_1=0$). As can be seen, isocurvature modes are strongly suppressed by $b_1 \ne 0$ in the super-Hubble regime compared to the case of standard kinetic terms: this effect is due to an effective mass for isocurvature perturbations which is larger than $H$ during the slow-roll regime. In the case of $k \sim k_\circ$, isocurvature perturbations grow for $b_1=4 / M_\mathrm{pl}$ and source in turn curvature ones. This effect is due to the fact that for high values of $b_1$, the effective mass-squared for isocurvature perturbations becomes temporarily negative during slow-roll violation, leading to a tachyonic growth of isocurvature.  
	
	We now turn to the primordial power spectra. Features in the power spectrum can be generated at the scales that cross the Hubble radius around the transition between the first and the second stage of inflation and depend on the combination of both $b_1$ and the ratio of the potentials $(m_\phi\,M_\textup{pl})^2 /V_0$.
	These are shown in Figs.~\ref{fig:power_Hubble} and \ref{fig:power}, where we plot the curvature, isocurvature and curvature-isocurvature cross-correlation power spectra at Hubble crossing and at the end of inflation, respectively, for the same parameters used in Fig.~\ref{fig:Background}.

	We plot a broad range of wavenumbers on the $x$-axis in units of $k_\circ$, that we define as the scale that crosses the Hubble radius when slow-roll is violated in the black dashed model\footnote{Note that $k_\circ$, although not very different, changes for every line in Fig.~\ref{fig:power} since we fix $N_*=50$ and different choices of $b_1$ give a different $N_\textup{end}$, as discussed earlier.} fixed by $b_1=2.4/ M_\textup{pl}$, i.e. at the peak of $\epsilon$ in the black dotted line of Fig.~\ref{fig:Background}. The red band is the range of scales that cross the Hubble radius during the break down of the slow-roll approximation. The part of the power spectrum that can be constrained by CMB observations consists of about $4$ e-folds  
	 and depends on the position of $k_\circ$ which, once fixed $N_*$, depends only on $\chi_i$.

	 As can be easily seen, apart from a small shift in the $k$-axis due to the different $k_\circ$, the power spectrum of curvature and isocurvature perturbations at the Hubble crossing is nearly the same for every value of $b_1$, as expected. 
	 The difference between curvature and isocurvature power spectra, however, is mainly due to the super-Hubble evolution of curvature and isocurvature perturbations, that we  discuss in the following.

Below we provide the features appearing at different cosmological scales:
	 \begin{itemize}

\item\textbf{Large scales}: 

We first look at the largest scales in Figs.~\ref{fig:power_Hubble} and \ref{fig:power}. If we decrease $\chi_i$ enough, slow-roll breaks down closer to the end of inflation and $k_\circ$ is shifted to the left of the plot. The power spectrum at CMB scales is then characterized by the spectral index predicted by chaotic inflation at the Hubble crossing. Note, however, that the amplitude of the scalar power spectrum depends on the non-canonical kinetic term $b_1$ because of the different isocurvature sourcing to $\mathcal{R}$.

	 	 The tensor power spectrum, as expected, is not affected by any sources as it is decoupled from the scalar perturbations at first order. This in turn leads to a lower tensor to scalar ratio as discussed above. We will discuss a specific example of this in the next section.

\item\textbf{Small scales}: 	 	 
	 
	 At smaller scales, 
	 we have the inflationary predictions of the single-field inflation driven by the field $\chi_i$, i.e. KKLT inflation at Hubble crossing. We have a small change in the curvature power spectrum at the end of inflation since its change induced by isocurvature perturbations is small. Isocurvature perturbations are indeed suppressed when their wavenumber exceed the Hubble radius and therefore acquire a blue spectrum in this region. All the spectra have a negligible dependence on $b_1$ in this region.
	 
	 The reason is that the field $\phi$ has already settled in the minimum of its potential $\phi_\textup{min}\simeq 0$ and thus the effects of $b(\phi_\textup{min})\simeq0$ are not as pronounced as in the first case.

	 The power spectrum of isocurvature perturbations in Fig.~\ref{fig:power} is indeed nearly the same for all the values of $b_1$ shown in the color-bar.

	 \item\textbf{Intermediate scales}:

This is the region where features are generated.
The wavenumbers in this region cross the Hubble
radius during the transition between the first and the second stage of inflation. 

As can be seen, since  the larger and the smaller scales have different amplitudes, the power spectrum has a sudden rise to match the two amplitudes. If $k_\circ\simeq 10^{-3} \,\textup{Mpc}^{-1}$  and the  plateau on the right side is properly normalized to the CMB normalization, then the junction between the two plateaus becomes a point of suppression of power at large scales. As we discuss in the next Section, this can be used to explain the low-$\ell$ deficit at $\ell \lesssim 40$ in the CMB temperature anisotropy pattern. 

Furthermore, as we increase $b_1$,  a bump, followed by a small, dip appears. This is easily explained by looking at the inset in top left panel of Fig.~\ref{fig:Background}, which contains an inset highlighting the evolution of the field $\phi$ near the transition. The field undergoes a damped oscillation around its minimum before settling in it. When $\phi$ decreases and eventually becomes negative, also $b(\phi)$ becomes negative and, if $b_1$ is large enough, the effective mass of isocurvature perturbations becomes negative 
as well for a few $e$-folds, leading to a brief instability for those modes that cross the Hubble radius around the violation of slow-roll.
 This results in an enhanced isocurvature feedback for the modes that cross the Hubble radius when $b(\phi)<0$. 
This effect is stronger for larger $b_1$ and the resulting bump might worsen the fit to CMB anisotropies. However, arbitrarily increasing  $b_1$ would further increase the bump: this mechanism might be used to generate Primordial Black Holes (PBH) and a stochastic background of Gravitational Waves (GW) from second order effects, if $k_\circ$ is pushed towards smaller scales.  This scenario is analyzed in Ref.~\cite{Braglia:2020eai}.

\end{itemize}

In the upper panel of Fig.~\ref{fig:power} we also illustrate the power spectrum of tensor modes. In this model, tensor modes are characterized by a larger amplitude for the modes which cross the Hubble radius during the first phase of inflation driven by a quadratic potential. The second phase of inflation is characterized by a tensor-to-scalar ratio typical of a KKLTI potential. This sort of broken power-law spectrum for tensor modes\footnote{See also Ref.~\cite{Pi:2019ihn} for a study of tensor perturbations in scenarios where inflation consists in two stages.} with a larger amplitude at large wavelengths might be therefore an interesting target for the next generation space missions dedicated to CMB polarization measurements as discussed in the next section.

\begin{figure}
	\begin{center} 
    		\resizebox{214pt}{172pt}{\includegraphics{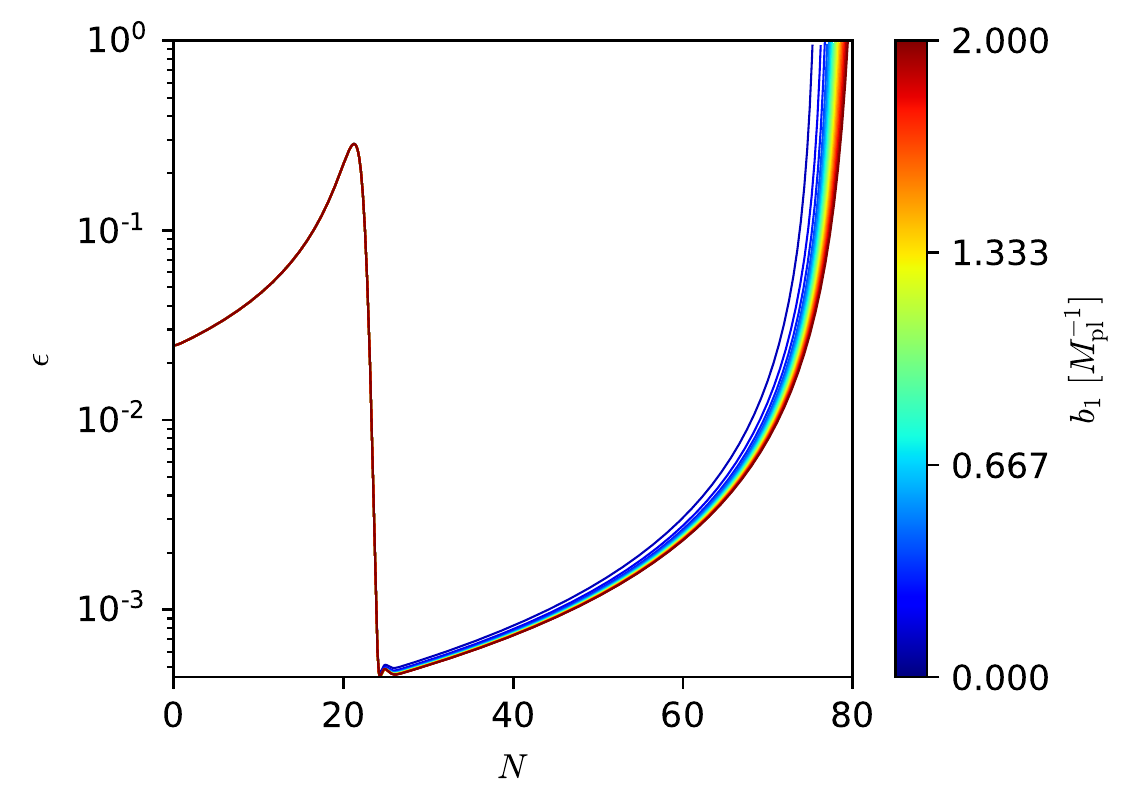}}			\resizebox{214pt}{172pt}{\includegraphics{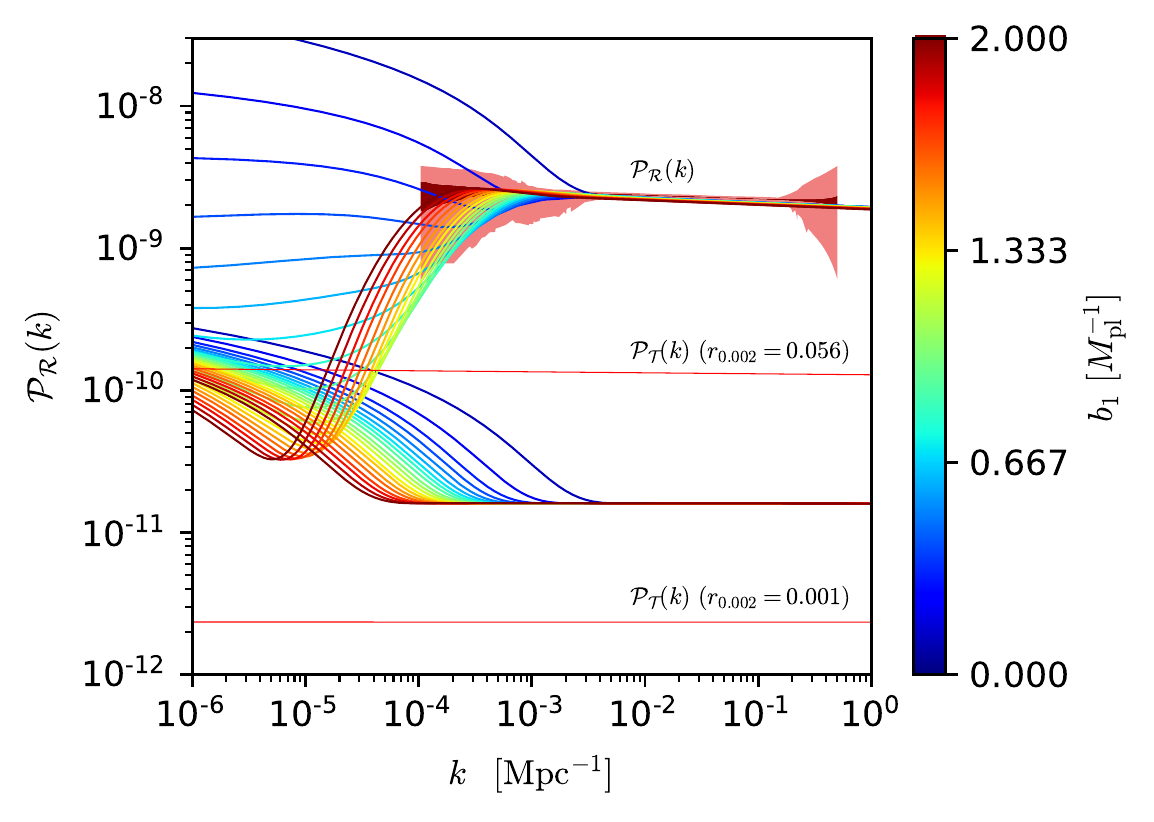}}		\resizebox{214pt}{172pt}{\includegraphics{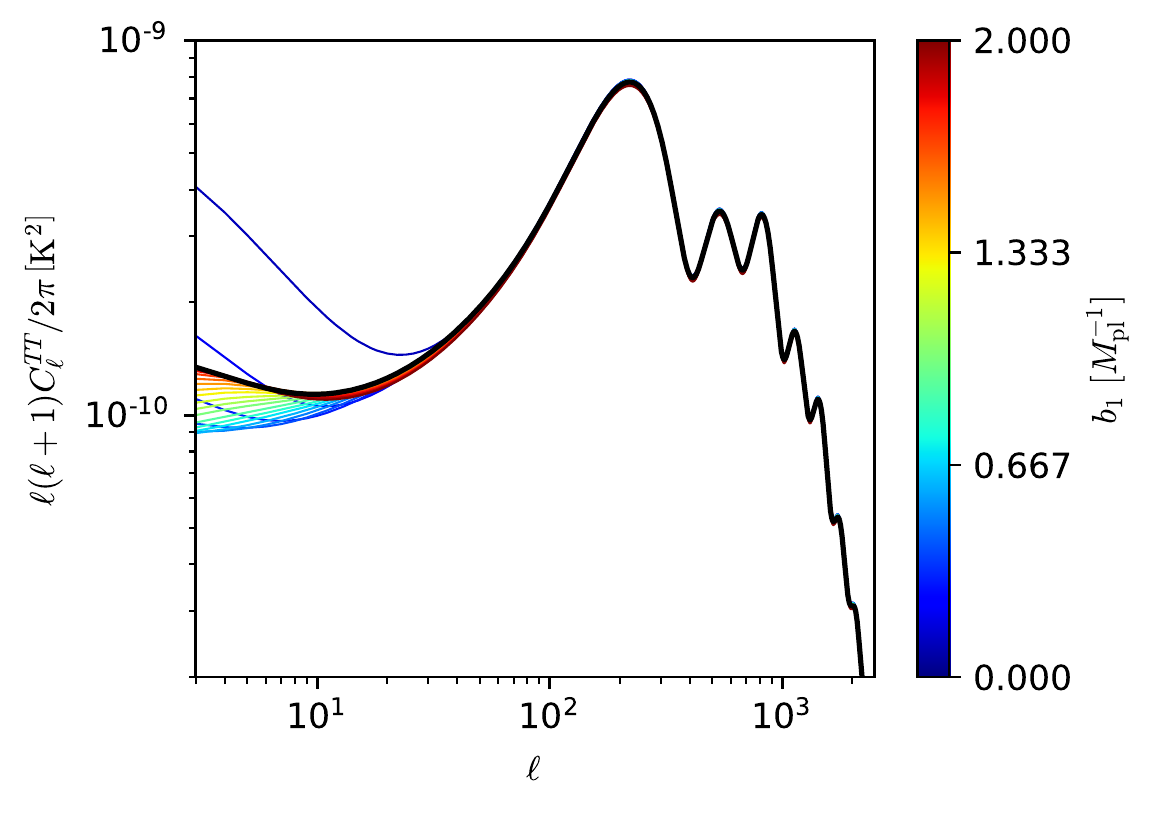}}
    		\resizebox{214pt}{172pt}{\includegraphics{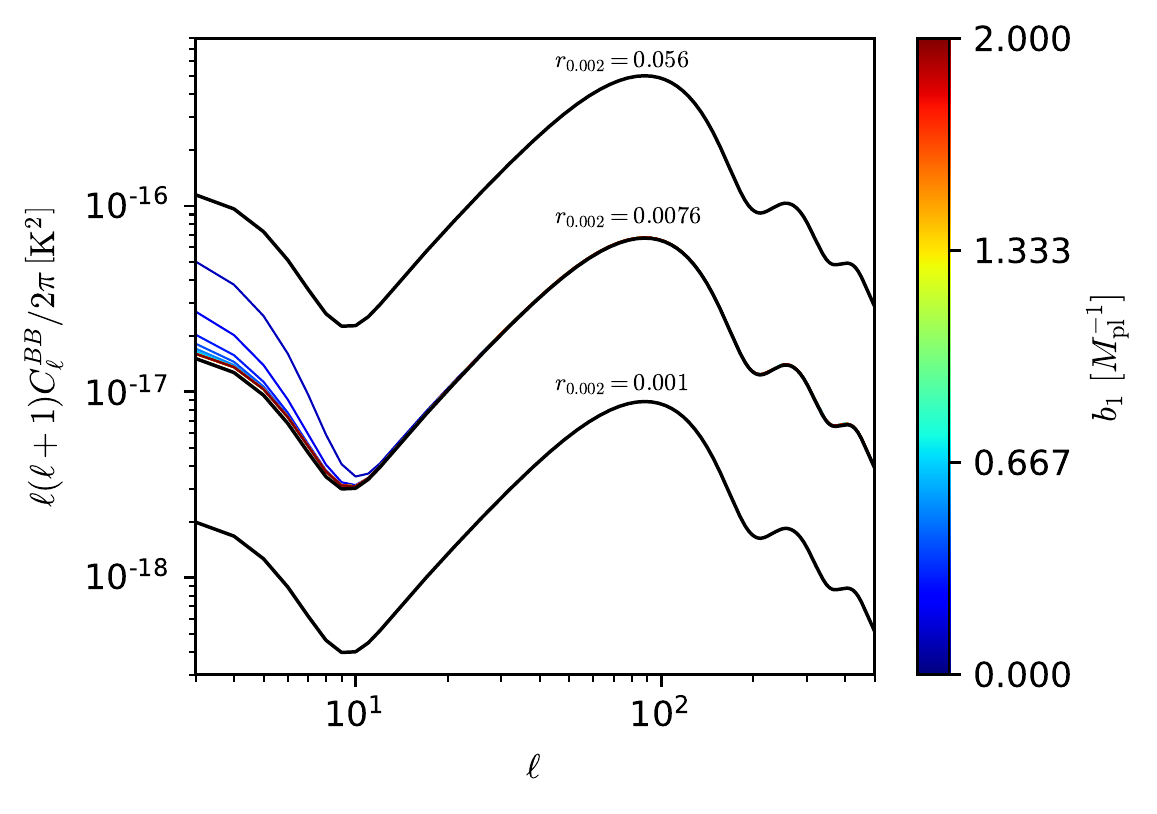}}	
	\end{center}
	\caption{\label{fig:Suppression}[Top-left]  $\epsilon$ parameter for a continuous range of values of $b_1$. [Top-right] curvature and  tensor  power spectra together with the $1$ and $2\sigma$ contours of the reconstructed primordial power spectrum by Planck \cite{Akrami:2018odb},  the upper bound on the tensor power spectrum (purple line) inferred by the combined analysis of  Planck, BAO, BICEP2, and 
		Keck Array \cite{Aghanim:2018eyx} and  the $r_{0.002}=0.001$ target of future CMB experiments \cite{Hazumi:2019lys,Delabrouille:2017rct,Hanany:2019wrm,Hanany:2019lle}. [Bottom-left] predictions on the CMB temperature lensed angular spectra. The black solid line is the Planck best-fit obtained with $r_{0.002}=0.056$. [Bottom-right] primordial B-mode angular spectra together with the ones computed with the consistency relation $r=-8 n_T$ for  $r_{0.002}=0.0056,\,0.0208,\,\text{and}\,0.001$ in black solid lines. The parameters used in the plot are 
	$(m_\phi\,M_\textup{pl})^2/V_0=1.07624$,  $\chi_0=\sqrt{3}M_\textup{pl}$ $\phi_i=8.8\,M_\textup{pl}$ and $\chi_i=5.76\,M_\textup{pl}$.}    
\end{figure}

\section{Effects on the CMB anisotropy power spectra}
	\label{sec:CMB}

We now show the importance of the non-canonical coupling in generating primordial features which could be connected to the deficit at large scales in the CMB temperature anisotropy power spectrum. In order to achieve a deficit at $\ell \lesssim 40$, we  shift the $k_\circ$ in Fig.~\ref{fig:power} around $k_\circ \sim 10^{-3}\,\textup{Mpc}^{-1}$. To do so, we need the second stage of inflation to last about $46-47$ $e$-folds (we remind that we fix $N_*=50$ in our analysis). This is achieved, by setting $\chi_i=5.76 M_\textup{pl}$. The evolution of $\epsilon$ for this model is given in the top left panel of Fig.~\ref{fig:Suppression}, where we have also listed all the parameters used in the numerical integration, that we have performed using a modified version of the publicly available code BINGO~\cite{Bingo}, that takes into account the full two fields dynamics. Note that the amplitude of the tensor power spectrum is larger than in Fig.~\ref{fig:power}, as we have used $\chi_0=\sqrt{3} \, M_\textup{pl}$, differently from Sec.~\ref{sec:results}.

In the top right panel, we plot the results for the curvature and tensor power spectra. We obtain a suppression of power on large scales starting from $0.7 M_\mathrm{pl}^{-1} \lesssim b_1 \lesssim 2 M_\mathrm{pl}^{-1} $. 
The tensor power spectrum shows a larger amplitude at low $k$, as also shown in Fig.~\ref{fig:power}, because the Hubble parameter $H$ is larger during the first stage of inflation.

The corresponding imprints on the CMB angular power spectra, that we computed  using the Einstein-Boltzmann code CLASS\footnote{\href{https://github.com/lesgourg/class\_public}{https://github.com/lesgourg/class\_public}} \cite{Blas:2011rf}, are shown in the lower panels of Fig.~\ref{fig:Suppression}. There is a lack of power in the CMB temperature power spectrum at low multipoles, without any modification to its small scale peak 
structure , more similar to what happens in Punctuated Inflation~\cite{Jain:2008dw} rather than for a discontinuity in the first derivative of the inflaton 
potential~\cite{Starobinsky:1994mh}. This mechanism can be therefore interesting to explain the low-$\ell$ deficit in the CMB temperature power spectrum in both the WMAP and Planck data \cite{Bennett2013,Aghanim:2018eyx} and is different from other ones such as Punctuated Inflation \cite{Jain:2008dw} or a discontinuity in the first derivative of the inflaton potential \cite{Starobinsky:1994mh}. For the same parameters, the larger amplitude in the tensor power spectrum at low $k$ induce a larger primordial B-mode polarization signal with respect to the single field realization of the second inflationary phase at $\ell \lesssim 30$, where the reionization bump is located. Again, this prediction for B-modes would be different from the corresponding one in presence of a discontinuity in the first derivative of the inflaton potential \cite{Starobinsky:1994mh}. We note, however, that in the model considered, all the larger primordial B-mode  occur for values of $b_1$ that give a higher contribution to the larger scale temperature spectra as well. 

\begin{figure}
	\begin{center} 
    		\resizebox{214pt}{172pt}{\includegraphics{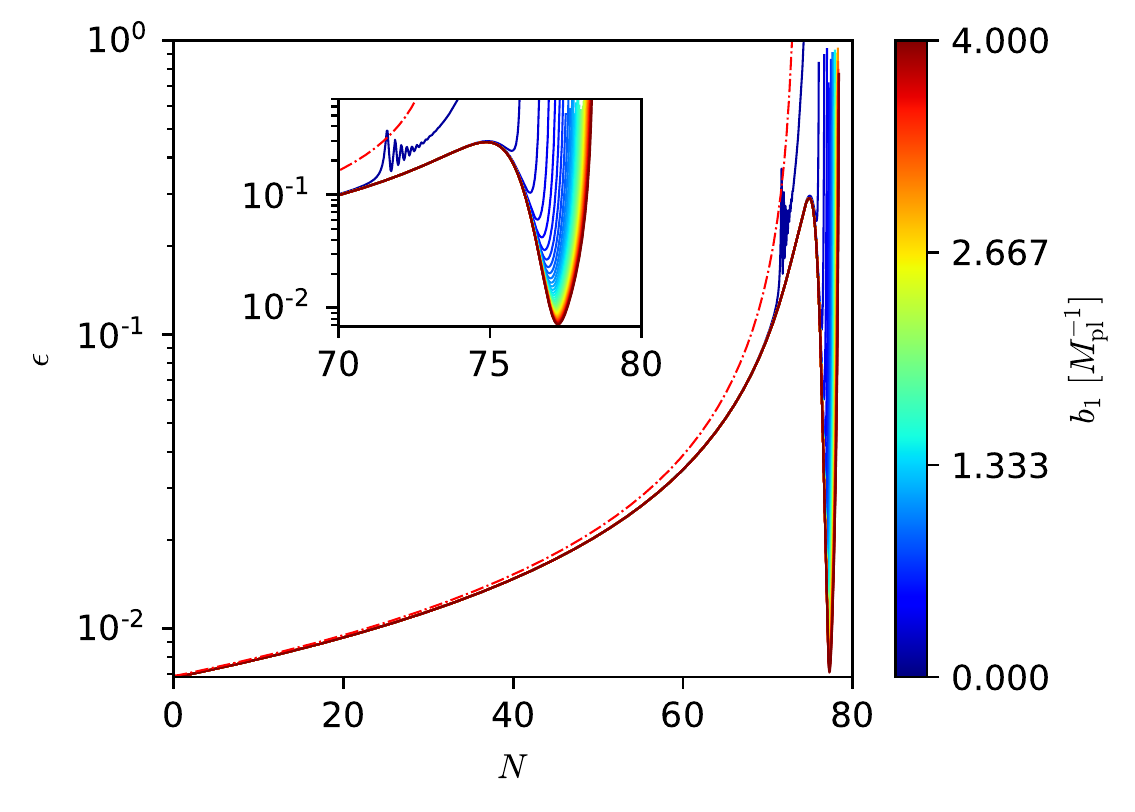}}		\resizebox{214pt}{172pt}{\includegraphics{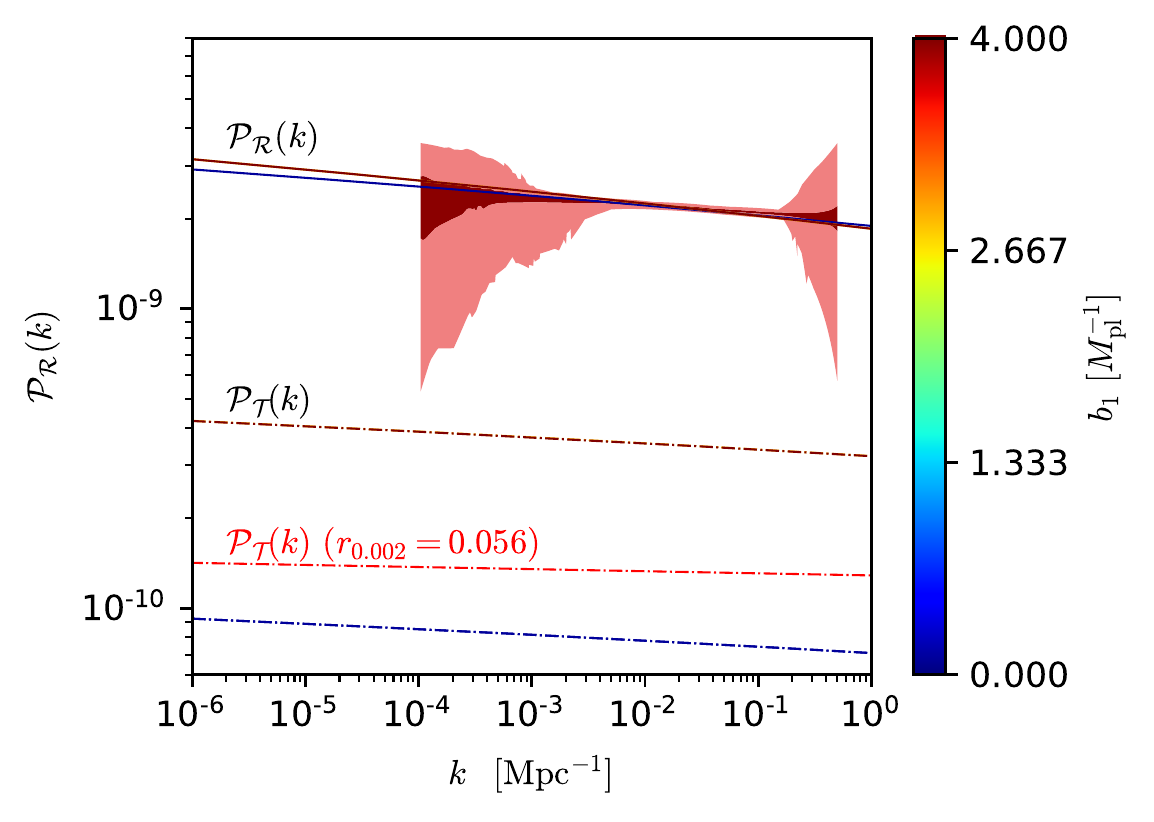}}		\resizebox{214pt}{172pt}{\includegraphics{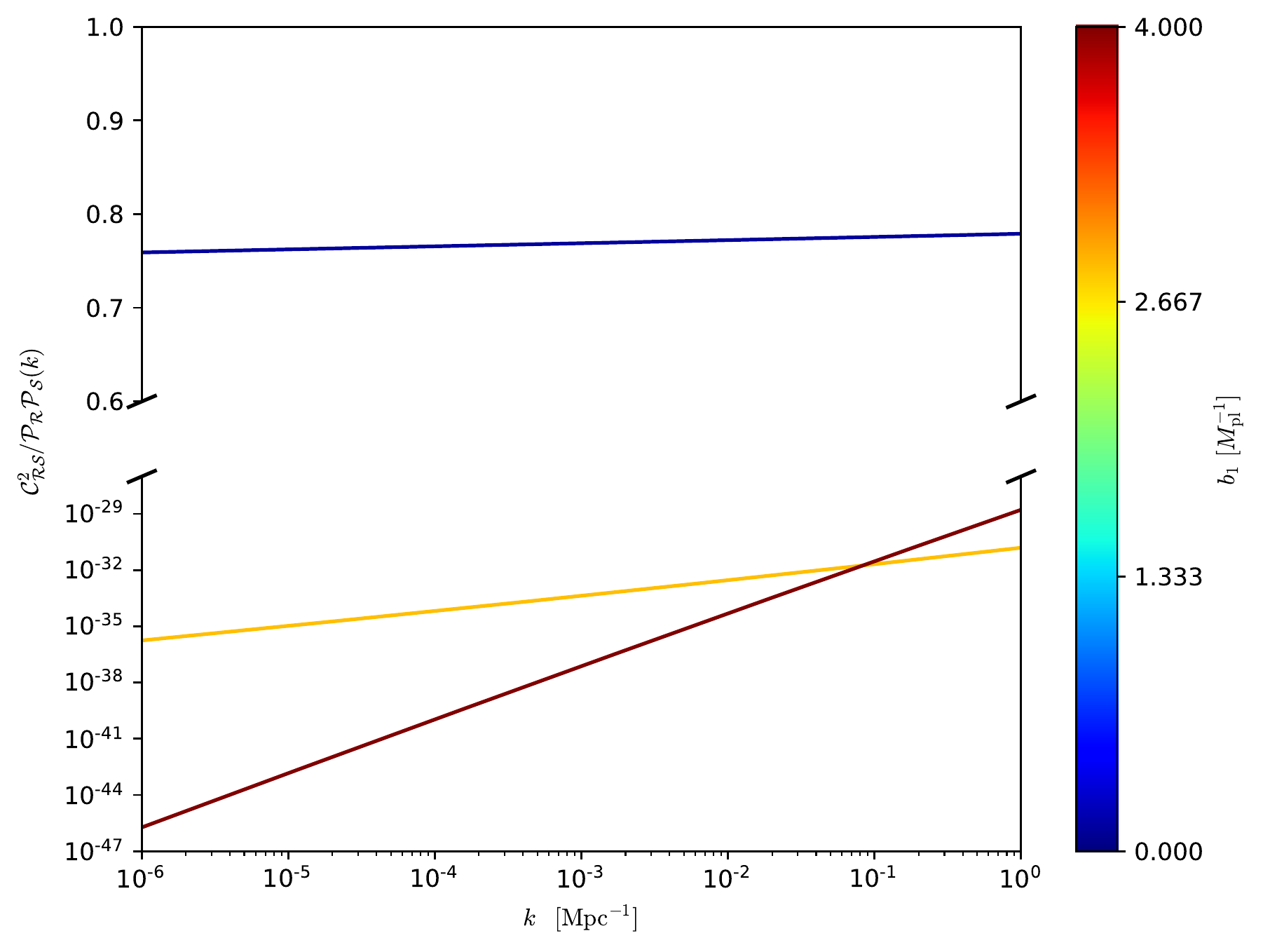}}
    		\resizebox{214pt}{172pt}{\includegraphics{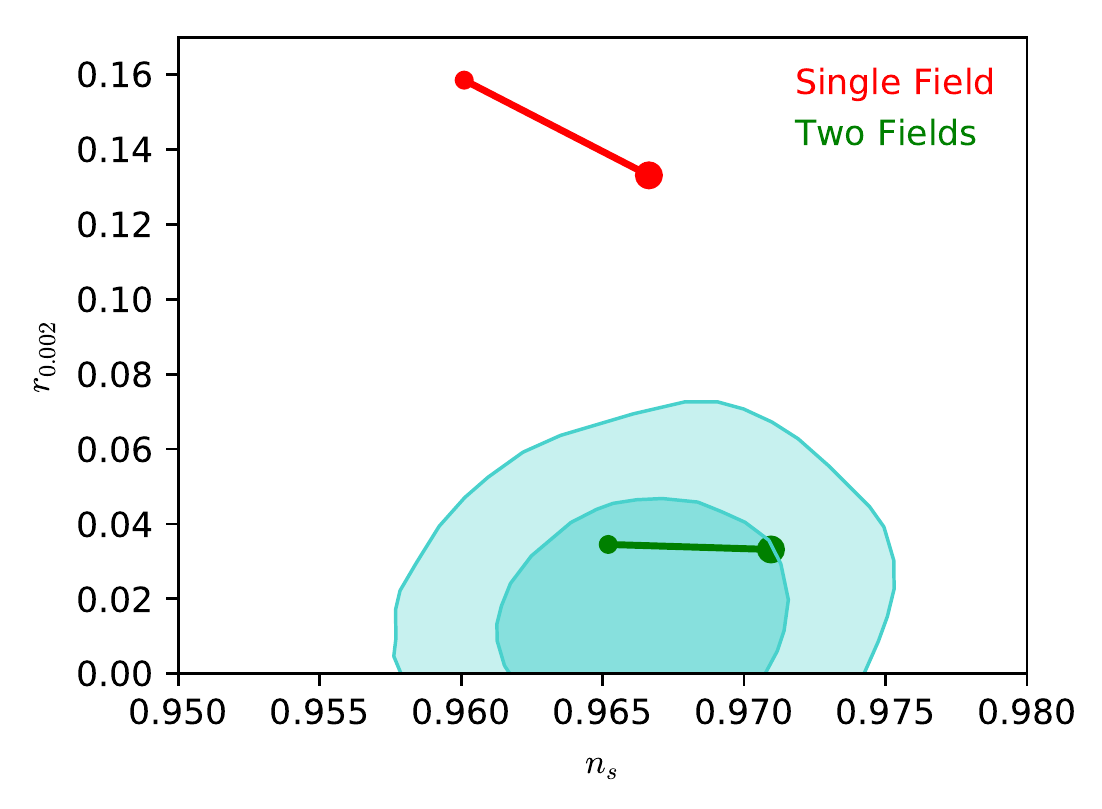}}	
	\end{center}
	\caption{\label{fig:LowTensor} [Top-left]  $\epsilon$ parameter. The red dashed line is the single-field version with the same $\phi_i$ and without the field $\chi$. [Top-right] curvature (solid), tensor (dashed) and isocurvature (dotted) power spectra for three models with $b_1=0,\,2.8,\,4.0$ (see color-bar) together with the upper bound on the tensor power spectrum (red line) inferred by the combined analysis of  Planck  and BICEP-Keck Array and the $1$ and $2\sigma$ contours of the reconstructed primordial power spectrum by Planck \cite{Akrami:2018odb}. Note that the ratio between the potential has been adjusted to match the amplitude $A_s$ of $\mathcal{P}_{\mathcal{R}}$. [Bottom-left] Curvature and isocurvature cross-correlation normalized to the curvature and isocurvature spectra. [Bottom-right] predictions in the $n_s-r$ plane of the chaotic single-field model (red) and our model with $b_1=0$ (green) on top of the $1$ and $2\sigma$ contours by Planck. Small (large) circles denote  $N_*=50\,(60)$. for a continuous range of values of $b_1$.  The parameters used in the plot are $(m_\phi\,M_\textup{pl})^2/V_0=1.17855$, $\phi_i=17\,M_\textup{pl}$ and $\chi_i=0.42 \,$ and in the three plots $b_1$ is varied for a continuous range of values.}    
\end{figure}

Let us now discuss the  
case in which the transition between the two stages occurs closer to the end of inflation, as shown in Fig.~\ref{fig:LowTensor}.
Whereas some differences in the scalar power spectrum and scalar tilt are present with respect to a single field realization of the first inflationary stage, i.e. a quadratic potential \cite{Linde:1983gd},  the tensor-to-scalar ratio can be suppressed by the correlation between curvature and isocurvature perturbations. The small differences in the scalar power spectra are due to the longer duration of inflation because of the second stage, as can be seen from the insert in top left panel of Fig.~\ref{fig:LowTensor}. Fig.~\ref{fig:LowTensor} also shows that, for the particular model in Eq.~\eqref{eq:potential}, the prediction for $r$ in the first inflationary phase driven by a quadratic potential can be reconciled with the most recent constraints by CMB anisotropies since cross-correlation between isocurvature and curvature perturbation can modify the relation between the tensor-to-scalar ratio and the tensor 
tilt~\cite{Bartolo:2001rt,Wands:2002bn,DiMarco:2005nq,Byrnes:2006fr,Dimopoulos:2005ac,Easther:2013rva,Price:2014ufa} 
\begin{equation}
r=-n_T\left(1-\frac{\mathcal{C}_{\mathcal{R}\mathcal{S}}^2}{\mathcal{P}_\mathcal{R}\mathcal{P}_\mathcal{S}}\right).
\end{equation}
As can be seen from the bottom left panel in Fig.~\ref{fig:LowTensor}, the cross-correlation is indeed larger when $b_1=0$, thus explaining the lower tensor-to-scalar ratio. We note that we have checked that $\mathcal{P}_\mathcal{S}$ is small for all the models in Fig.~\ref{fig:LowTensor}.

Note that the mechanism just presented differs from earlier attempts to cure  single field models that are in tension with the Planck data with the addition of heavy \cite{Achucarro:2015rfa,Renaux-Petel:2015mga} and/or spectator fields during inflation \cite{Langlois:2004nn,Moroi:2005kz,Moroi:2005np,Ichikawa:2008iq,Enqvist:2013paa,Vennin:2015vfa}. In fact, it is the presence of a second, lighter inflaton that modifies the single-field predictions in our model.

	\section{Discussions and conclusions}
	\label{sec:discussion}
	
Generating features at large scales in single-field inflationary models requires in general either a fine-tuned scalar field Lagrangian or exotic initial conditions in the background dynamics/quantum fluctuations. 
On the other hand, generating  features of similar shape by turning trajectories in field space in multi-field inflationary models requires also a careful study of isocurvature modes and their effects on the curvature perturbations. 
Isocurvature perturbations can indeed erase features in the PPS generated at Hubble crossing, as we have explicitly shown for the archetypal case of inflation driven by two massive scalar fields~\cite{Polarski:1994rz}, leading to different conclusions from~\cite{Feng:2003zua}.
Isocurvature fluctuations can have an impact on other inspired multi-field models for primordial features as well.

We have then discussed the phenomenology of features at large scales in two field inflationary models in which there is 
a non-trivial kinetic term of a second field~$\chi$ coupled to~$\phi$. 
We have restricted ourselves to the case of a separable potential in which features can be generated during a breakdown of the slow-roll regime between two phases of inflation driven first by a scalar field with a larger effective mass and then 
by a second field with a smaller one.
We have numerically integrated the exact equations for the linear perturbations taking into account the super-Hubble evolution of curvature and  isocurvature perturbations. 

We have shown how a sufficiently large coupling ($b \gtrsim M^{-1}_\mathrm{pl}$) with the kinetic term of the second field can lead to a suppression of isocurvature perturbations, therefore decreasing their feedback into curvature perturbations. We have computed the associated CMB power spectra showing the phenomenological relevance of this mechanism to produce the low-$\ell$ deficit observed in the temperature power spectrum. For the same parameters, we have also shown how the tensor power spectrum can have a larger amplitude at small $k$ since the Hubble parameter is larger in the first inflationary phase. 
For this last reason, the mechanism discussed here is potentially different from other single field inflationary models which produce a deficit in the CMB temperature power spectrum at low multipoles, without decreasing the B-mode polarization signal as for a short inflationary stage preceeded by a kinetic stage \cite{Contaldi:2003zv,Nicholson:2007by}.
This possibility is obviously of interest for the next generation experiments dedicated to CMB polarization which can have access to low multipoles \cite{Hazumi:2019lys,Delabrouille:2017rct,Hanany:2019wrm}. 

On the other hand, we have also shown that, when the breakdown of slow-roll occurs close to the end of inflation, isocurvature perturbations can enhance the scalar PPS without altering the tensor one if the two fields are canonically coupled, and therefore reducing the tensor-to-scalar ratio on CMB scales.

	In terms of future work, we note that the breakdown of the slow-roll regime would probably lead to a certain level of non-Gaussianity.  
	We hope to study this issue in the future.
	
	\section*{Acknowledgements}
DKH has received fundings from the European Union’s Horizon 2020 research and innovation programme under the Marie Sklodowska-Curie grant agreement  No. 664931.
FF acknowledges financial 
support by ASI Grant 2016-24-H.0.
LS wishes to acknowledge support from the Science and Engineering Research  
Board, Department of Science and Technology, Government of India, through 
the Core Research Grant CRG/2018/002200.

	\appendix
	\section{Varying the ratio of the potentials}\label{appendix:VaryingMass}	
        	\begin{figure}
    	\begin{center} 
    		\resizebox{214pt}{172pt}{\includegraphics{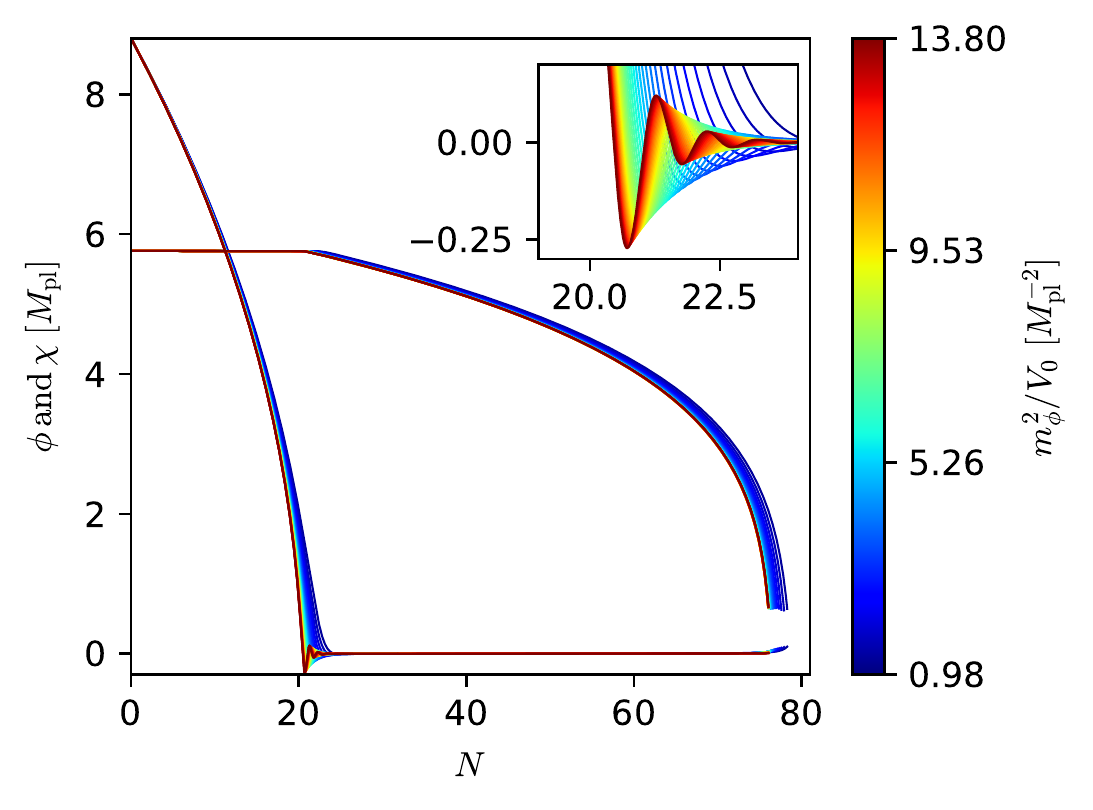}}			\resizebox{214pt}{172pt}{\includegraphics{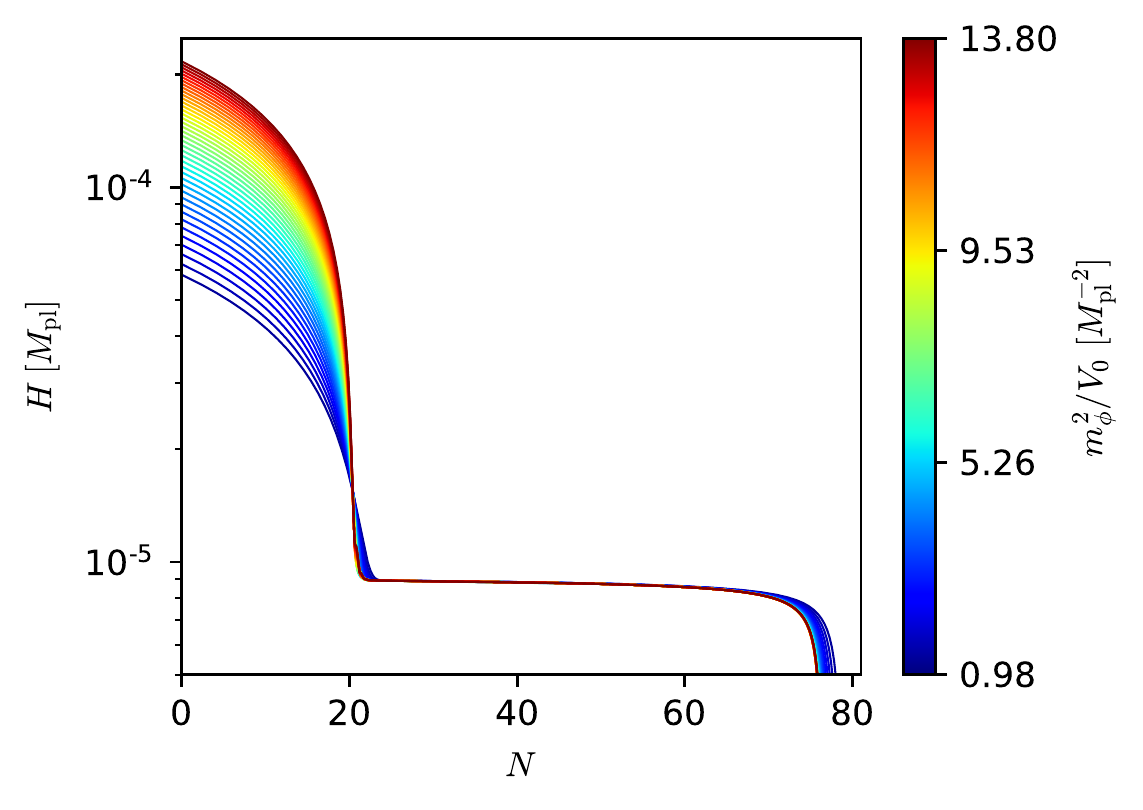}}		\resizebox{214pt}{172pt}{\includegraphics{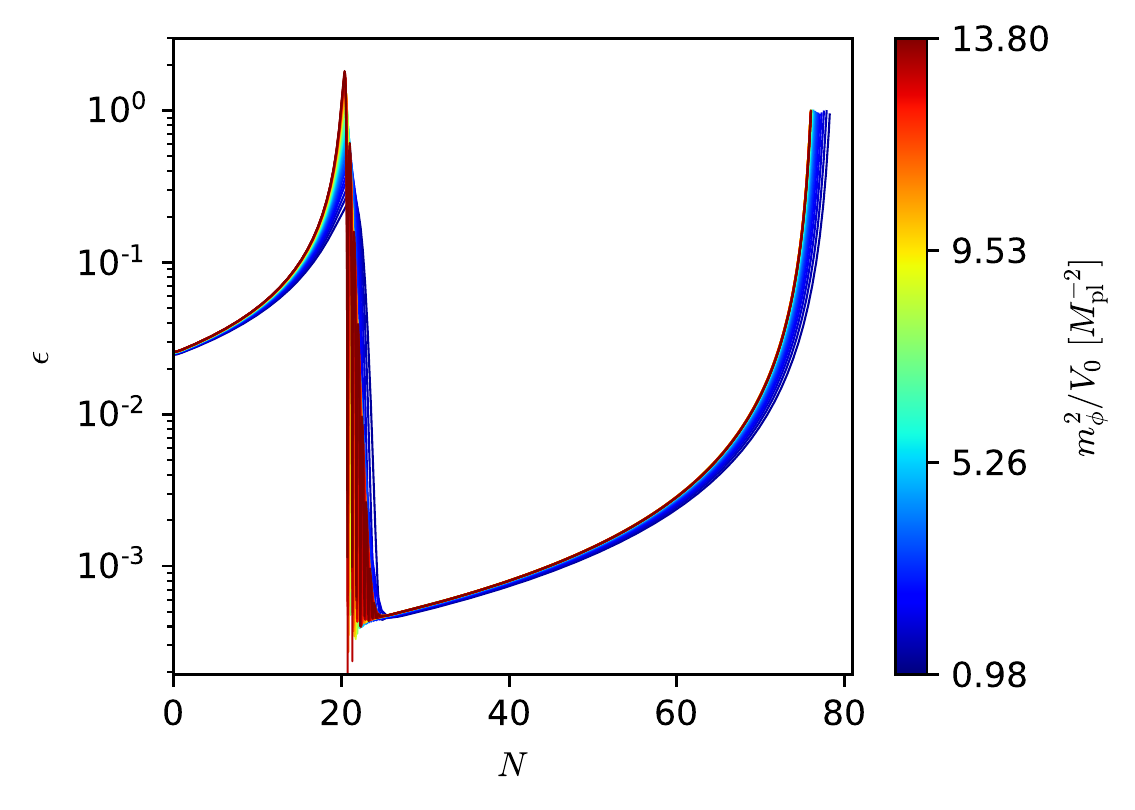}}
    		\resizebox{214pt}{172pt}{\includegraphics{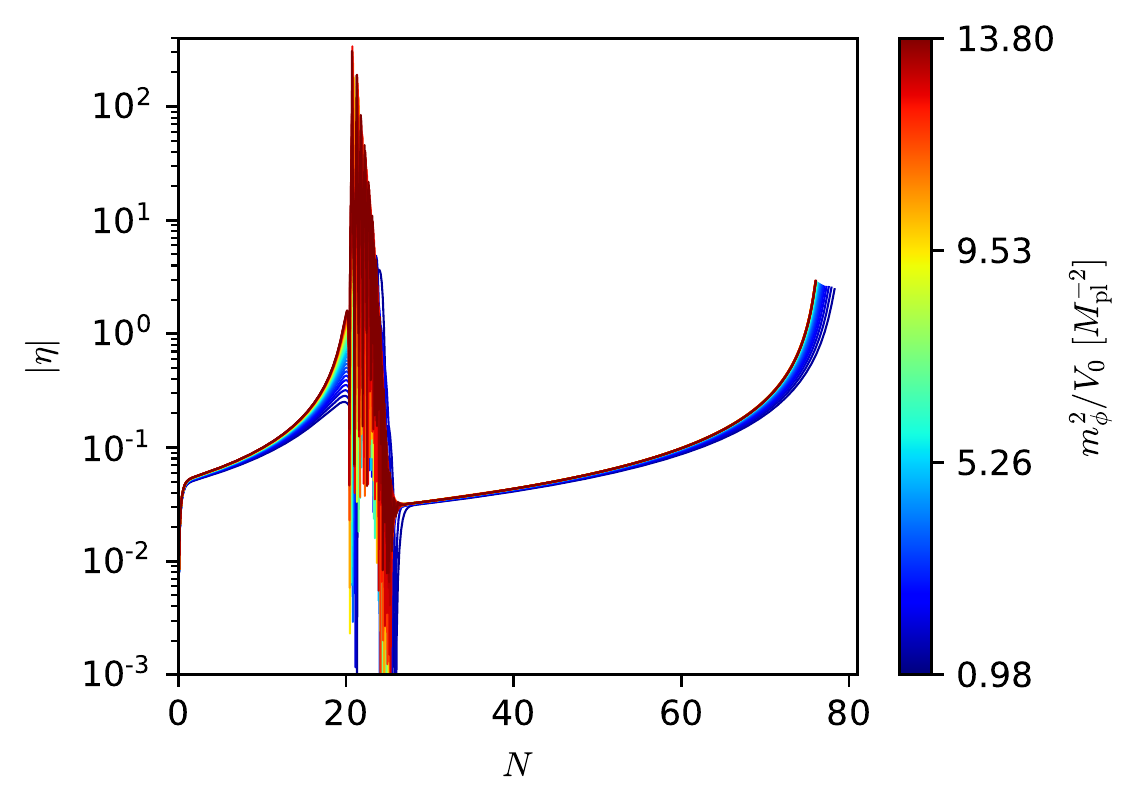}}	
    	\end{center}
    	\caption{\label{fig:Background2}[Top-left] Scalar fields and [top-right]  Hubble parameter $H$  evolution. [Bottom-left] First slow-roll parameter $\epsilon$   and [top-right] second slow-roll parameter $\eta\equiv\epsilon_2$.  The non-canonical coupling $b_1=1/M_\textup{pl}$ is fixed   and we vary the potential ratio $(m_\phi\,M_\textup{pl})^2 /V_0$  for a continuous range of values.} 
    \end{figure}

\begin{figure}
	\begin{center} 
		\resizebox{214pt}{172pt}{\includegraphics{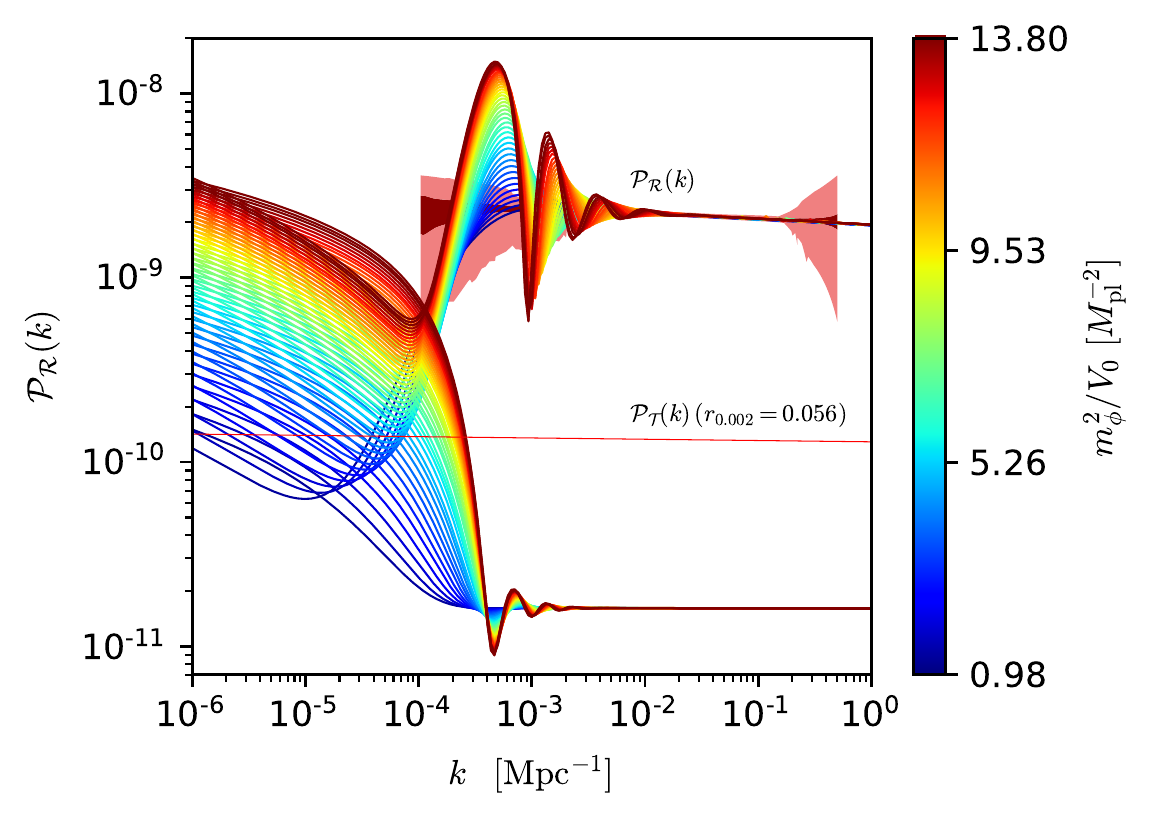}}			\resizebox{214pt}{172pt}{\includegraphics{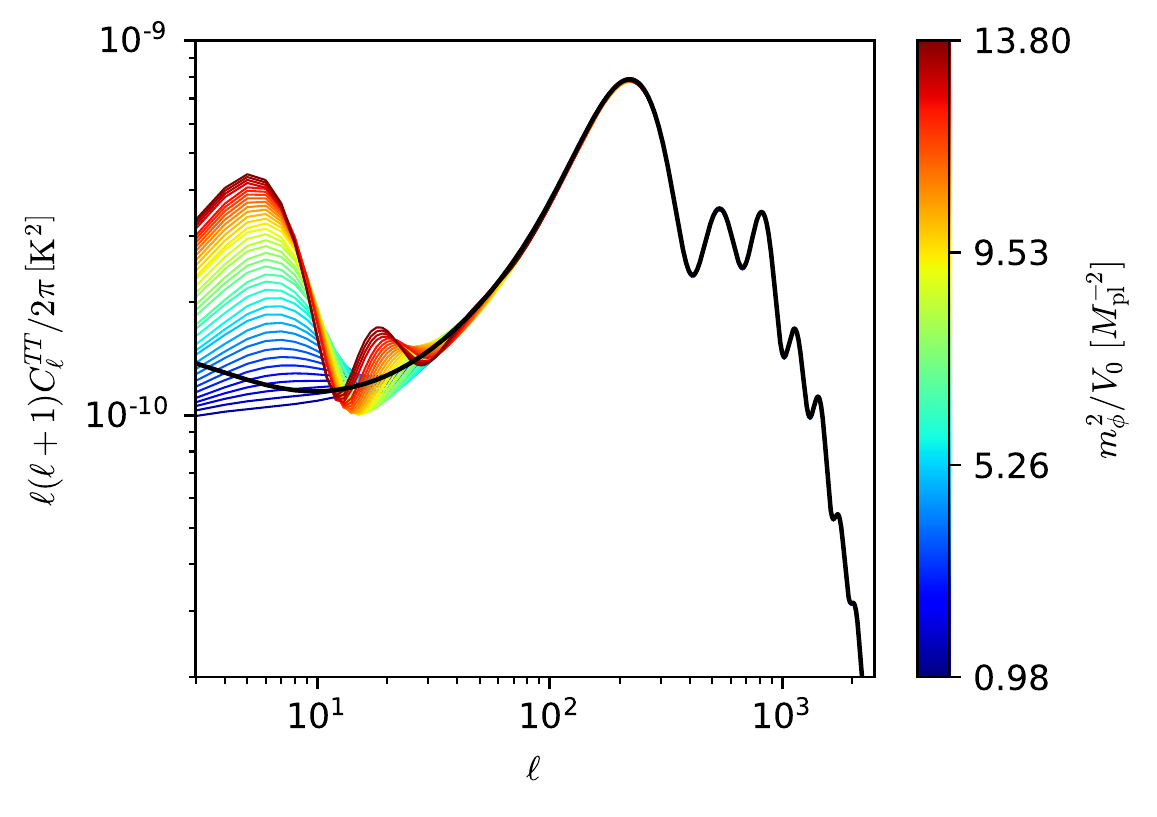}}	
	\end{center}
	\caption{\label{fig:powerMass}[Left] scalar and tensor power spectrum  for the model Eq.~\eqref{eq:potential}.   [Right] predictions on the CMB temperature lensed angular spectra. The red dashed line denotes the Planck 18 + BK15 95 \% CL upper bound on $r$, i.e. $r_{0.002}=0.056$. The non-canonical coupling $b_1=1/M_\textup{pl}$ is fixed   and we vary the potential ratio $(m_\phi\,M_\textup{pl})^2 /V_0$  for a continuous range of values.} 
\end{figure}

    In the main text, we have shown the effects of the non-canonical kinetic coupling of the two scalar fields. In particular, we have observed that, at a fixed potential ratio, higher values of the coupling $b_1$ can either enhance or suppress isocurvature perturbations and their feedback into the scalar PPS, depending on the sign of the scalar field $\phi$. The goal of this Appendix is to provide a full scan of the parameter space of our toy model, showing the effect of varying the potential ratio $(m_\phi\,M_\textup{pl})^2 /V_0$. 
    
    The background evolution of the model is shown in Fig.~\ref{fig:Background2}. As can be seen, by increasing the ratio of the potentials, the energy scale of the first inflationary stage becomes higher and the transition between the two stages more violent. This results in a huge deviation from $\eta\simeq \mathcal{O}(1)$ of the second slow-roll parameter. Also, the first slow-roll parameter can become even greater than unity\footnote{We mention that such a huge deviation would probably lead to large non-Gaussianities. However, the non-Gaussian nature of the perturbations is not the purpose of this paper.}, leading to  an intermediate matter-dominated expansion~\cite{Polarski:1992dq}. During this transition, the heavier scalar field undergoes damped oscillations around its minimum. These can last up to $\sim5$ $e$-folds as can be seen from the  inset in the top left panel of Fig.~\ref{fig:Background2}.
    
    Following the reasoning of Section~\ref{sec:results}, the damped oscillations between positive and negative values of the field $\phi$ leads to a continuous suppression and increase of  isocurvature sourcing to the curvature perturbation. This series of peaks and dips can be thus seen in the power spectrum at  scales that leave the horizon during this background oscillatory phase. This is shown in the left panel of Fig.~\ref{fig:powerMass}, where we plot the scalar and tensor power spectrum for our model  obtained by fixing $b_1$ and varying the potential ratio. As can be seen, the oscillating pattern of the field $\phi$ is indeed imprinted in both the scalar and tensor power spectra. 
    
    Nevertheless, the damped oscillations have a high amplitude at large scales. The oscillations have therefore to be at unobservable scales to fit the CMB data.   In this case the CMB angular power spectra are in fact very similar to those obtained using a power-law power spectrum. On the other hand,  if the oscillations in the power spectrum are in the  range $10^{-4}-10^{-2}$ $\text{Mpc}^{-1}$ as in the left panel of Fig.~\ref{fig:powerMass}, we obtain the CMB angular power spectra in the right panel of Fig.~\ref{fig:powerMass}. As can be seen, the range of multipoles $\ell\leq 100$ is totally different from the $\Lambda$CDM best-fit plotted in solid black lines and the fit to data is considerably affected.

	%%%%%%%%%%%%%%%%%%%%%%%%%%%%%%%%%%%%%%%%%%%%%%%%%%%%%%%%%%%%%%%%%%%%%%%%%%%%%%%
	\bibliographystyle{JHEP}
	\bibliography{TFF}

\providecommand{\href}[2]{#2}\begingroup\raggedright\begin{thebibliography}{100}

\bibitem{Akrami:2018odb}
{\scshape Planck} collaboration, Y.~Akrami et~al., \emph{{Planck 2018 results.
  X. Constraints on inflation}},
  \href{https://arxiv.org/abs/1807.06211}{{\ttfamily 1807.06211}}.

\bibitem{Akrami:2019izv}
{\scshape Planck} collaboration, Y.~Akrami et~al., \emph{{Planck 2018 results.
  IX. Constraints on primordial non-Gaussianity}},
  \href{https://arxiv.org/abs/1905.05697}{{\ttfamily 1905.05697}}.

\bibitem{Hannestad2001}
S.~Hannestad, \emph{Reconstructing the inflationary power spectrum from cosmic
  microwave background radiation data},
  \href{https://doi.org/10.1103/physrevd.63.043009}{\emph{Physical Review D}
  {\bfseries 63} (2001) }.

\bibitem{Tegmark2002}
M.~Tegmark and M.~Zaldarriaga, \emph{Separating the early universe from the
  late universe: Cosmological parameter estimation beyond the black box},
  \href{https://doi.org/10.1103/physrevd.66.103508}{\emph{Physical Review D}
  {\bfseries 66} (2002) }.

\bibitem{Mukherjee2003}
P.~Mukherjee and Y.~Wang, \emph{Model‐independent reconstruction of the
  primordial power spectrum fromwilkinson microwave anistropy probedata},
  \href{https://doi.org/10.1086/379161}{\emph{The Astrophysical Journal}
  {\bfseries 599} (2003) 1}.

\bibitem{TocchiniValentini:2005ja}
D.~Tocchini-Valentini, Y.~Hoffman and J.~Silk, \emph{{Non-parametric
  reconstruction of the primordial power spectrum at horizon scales from wmap
  data}}, \href{https://doi.org/10.1111/j.1365-2966.2006.10031.x}{\emph{Mon.
  Not. Roy. Astron. Soc.} {\bfseries 367} (2006) 1095}
  [\href{https://arxiv.org/abs/astro-ph/0509478}{{\ttfamily
  astro-ph/0509478}}].

\bibitem{Shafieloo2004}
A.~Shafieloo and T.~Souradeep, \emph{Primordial power spectrum from wmap},
  \href{https://doi.org/10.1103/physrevd.70.043523}{\emph{Physical Review D}
  {\bfseries 70} (2004) }.

\bibitem{Kogo2005}
N.~Kogo, M.~Sasaki and J.~Yokoyama, \emph{Constraining cosmological parameters
  by the cosmic inversion method},
  \href{https://doi.org/10.1143/ptp.114.555}{\emph{Progress of Theoretical
  Physics} {\bfseries 114} (2005) 555}.

\bibitem{Leach:2005av}
S.~M. Leach, \emph{{Measuring the primordial power spectrum: Principal
  component analysis of the cosmic microwave background}},
  \href{https://doi.org/10.1111/j.1365-2966.2006.10842.x}{\emph{Mon. Not. Roy.
  Astron. Soc.} {\bfseries 372} (2006) 646}
  [\href{https://arxiv.org/abs/astro-ph/0506390}{{\ttfamily
  astro-ph/0506390}}].

\bibitem{Shafieloo2008}
A.~Shafieloo and T.~Souradeep, \emph{Estimation of primordial spectrum with
  post-wmap 3-year data},
  \href{https://doi.org/10.1103/physrevd.78.023511}{\emph{Physical Review D}
  {\bfseries 78} (2008) }.

\bibitem{Paykari2010}
P.~Paykari and A.~H. Jaffe, \emph{Optimal binning of the primordial power
  spectrum}, \href{https://doi.org/10.1088/0004-637x/711/1/1}{\emph{The
  Astrophysical Journal} {\bfseries 711} (2010) 1}.

\bibitem{Nicholson:2009pi}
G.~Nicholson and C.~R. Contaldi, \emph{{Reconstruction of the Primordial Power
  Spectrum using Temperature and Polarisation Data from Multiple Experiments}},
  \href{https://doi.org/10.1088/1475-7516/2009/07/011}{\emph{JCAP} {\bfseries
  07} (2009) 011} [\href{https://arxiv.org/abs/0903.1106}{{\ttfamily
  0903.1106}}].

\bibitem{Gauthier2012}
C.~Gauthier and M.~Bucher, \emph{Reconstructing the primordial power spectrum
  from the cmb},
  \href{https://doi.org/10.1088/1475-7516/2012/10/050}{\emph{Journal of
  Cosmology and Astroparticle Physics} {\bfseries 2012} (2012) 050}.

\bibitem{Hlozek2012}
R.~Hlozek, J.~Dunkley, G.~Addison, J.~W. Appel, J.~R. Bond, C.~S. Carvalho
  et~al., \emph{The atacama cosmology telescope: A measurement of the
  primordial power spectrum},
  \href{https://doi.org/10.1088/0004-637x/749/1/90}{\emph{The Astrophysical
  Journal} {\bfseries 749} (2012) 90}.

\bibitem{Vzquez2012}
J.~A. V{\'a}zquez, M.~Bridges, M.~Hobson and A.~Lasenby, \emph{Model selection
  applied to reconstruction of the primordial power spectrum},
  \href{https://doi.org/10.1088/1475-7516/2012/06/006}{\emph{Journal of
  Cosmology and Astroparticle Physics} {\bfseries 2012} (2012) 006}.

\bibitem{Hazra2013}
D.~K. Hazra, A.~Shafieloo and T.~Souradeep, \emph{Primordial power spectrum: a
  complete analysis with the wmap nine-year data},
  \href{https://doi.org/10.1088/1475-7516/2013/07/031}{\emph{Journal of
  Cosmology and Astroparticle Physics} {\bfseries 2013} (2013) 031}.

\bibitem{Hunt2014}
P.~Hunt and S.~Sarkar, \emph{Reconstruction of the primordial power spectrum of
  curvature perturbations using multiple data sets},
  \href{https://doi.org/10.1088/1475-7516/2014/01/025}{\emph{Journal of
  Cosmology and Astroparticle Physics} {\bfseries 2014} (2014) 025}.

\bibitem{Hazra:2014jwa}
D.~K. Hazra, A.~Shafieloo and T.~Souradeep, \emph{{Primordial power spectrum
  from Planck}},
  \href{https://doi.org/10.1088/1475-7516/2014/11/011}{\emph{JCAP} {\bfseries
  1411} (2014) 011} [\href{https://arxiv.org/abs/1406.4827}{{\ttfamily
  1406.4827}}].

\bibitem{Brando:2020yvo}
G.~Brando and E.~V. Linder, \emph{{Exploring Early and Late Cosmology with Next
  Generation Surveys}},  \href{https://arxiv.org/abs/2001.07738}{{\ttfamily
  2001.07738}}.

\bibitem{Starobinsky:1992ts}
A.~A. Starobinsky, \emph{{Spectrum of adiabatic perturbations in the universe
  when there are singularities in the inflation potential}}, {\emph{JETP Lett.}
  {\bfseries 55} (1992) 489}.

\bibitem{Contaldi:2003zv}
C.~R. Contaldi, M.~Peloso, L.~Kofman and A.~D. Linde, \emph{{Suppressing the
  lower multipoles in the CMB anisotropies}},
  \href{https://doi.org/10.1088/1475-7516/2003/07/002}{\emph{JCAP} {\bfseries
  0307} (2003) 002} [\href{https://arxiv.org/abs/astro-ph/0303636}{{\ttfamily
  astro-ph/0303636}}].

\bibitem{Sriramkumar:2004pj}
L.~Sriramkumar and T.~Padmanabhan, \emph{{Initial state of matter fields and
  trans-Planckian physics: Can CMB observations disentangle the two?}},
  \href{https://doi.org/10.1103/PhysRevD.71.103512}{\emph{Phys. Rev.}
  {\bfseries D71} (2005) 103512}
  [\href{https://arxiv.org/abs/gr-qc/0408034}{{\ttfamily gr-qc/0408034}}].

\bibitem{Allahverdi:2006wt}
R.~{Allahverdi} and A.~{Mazumdar}, \emph{{Spectral tilt in A-term inflation}},
  {\emph{arXiv e-prints} (2006) hep}
  [\href{https://arxiv.org/abs/hep-ph/0610069}{{\ttfamily hep-ph/0610069}}].

\bibitem{Jain:2007au}
R.~K. Jain, P.~Chingangbam and L.~Sriramkumar, \emph{{On the evolution of
  tachyonic perturbations at super-Hubble scales}},
  \href{https://doi.org/10.1088/1475-7516/2007/10/003}{\emph{JCAP} {\bfseries
  10} (2007) 003} [\href{https://arxiv.org/abs/astro-ph/0703762}{{\ttfamily
  astro-ph/0703762}}].

\bibitem{Jain:2008dw}
R.~K. Jain, P.~Chingangbam, J.-O. Gong, L.~Sriramkumar and T.~Souradeep,
  \emph{{Punctuated inflation and the low CMB multipoles}},
  \href{https://doi.org/10.1088/1475-7516/2009/01/009}{\emph{JCAP} {\bfseries
  0901} (2009) 009} [\href{https://arxiv.org/abs/0809.3915}{{\ttfamily
  0809.3915}}].

\bibitem{Jain:2009pm}
R.~K. Jain, P.~Chingangbam, L.~Sriramkumar and T.~Souradeep, \emph{{The
  tensor-to-scalar ratio in punctuated inflation}},
  \href{https://doi.org/10.1103/PhysRevD.82.023509}{\emph{Phys.\ Rev.\ D}
  {\bfseries 82} (2010) 023509}
  [\href{https://arxiv.org/abs/0904.2518}{{\ttfamily 0904.2518}}].

\bibitem{Ramirez:2011kk}
E.~Ramirez and D.~J. Schwarz, \emph{{Predictions of just-enough inflation}},
  \href{https://doi.org/10.1103/PhysRevD.85.103516}{\emph{Phys.\ Rev.\ D}
  {\bfseries 85} (2012) 103516}
  [\href{https://arxiv.org/abs/1111.7131}{{\ttfamily 1111.7131}}].

\bibitem{Ramirez:2012gt}
E.~Ramirez, \emph{{Low power on large scales in just enough inflation models}},
  \href{https://doi.org/10.1103/PhysRevD.85.103517}{\emph{Phys.\ Rev.\ D}
  {\bfseries 85} (2012) 103517}
  [\href{https://arxiv.org/abs/1202.0698}{{\ttfamily 1202.0698}}].

\bibitem{Hazra:2014goa}
D.~K. Hazra, A.~Shafieloo, G.~F. Smoot and A.~A. Starobinsky, \emph{{Wiggly
  Whipped Inflation}},
  \href{https://doi.org/10.1088/1475-7516/2014/08/048}{\emph{JCAP} {\bfseries
  08} (2014) 048} [\href{https://arxiv.org/abs/1405.2012}{{\ttfamily
  1405.2012}}].

\bibitem{Hazra2014b}
D.~K. {Hazra}, A.~{Shafieloo}, G.~F. {Smoot} and A.~A. {Starobinsky},
  \emph{{Inflation with Whip-Shaped Suppressed Scalar Power Spectra}},
  \href{https://doi.org/10.1103/PhysRevLett.113.071301}{\emph{Physical Review
  Letters} {\bfseries 113} (2014) 071301}
  [\href{https://arxiv.org/abs/1404.0360}{{\ttfamily 1404.0360}}].

\bibitem{Bousso:2014jca}
R.~Bousso, D.~Harlow and L.~Senatore, \emph{{Inflation After False Vacuum
  Decay: New Evidence from BICEP2}},
  \href{https://doi.org/10.1088/1475-7516/2014/12/019}{\emph{JCAP} {\bfseries
  1412} (2014) 019} [\href{https://arxiv.org/abs/1404.2278}{{\ttfamily
  1404.2278}}].

\bibitem{Hazra:2016fkm}
D.~K. Hazra, A.~Shafieloo, G.~F. Smoot and A.~A. Starobinsky, \emph{{Primordial
  features and Planck polarization}},
  \href{https://doi.org/10.1088/1475-7516/2016/09/009}{\emph{JCAP} {\bfseries
  1609} (2016) 009} [\href{https://arxiv.org/abs/1605.02106}{{\ttfamily
  1605.02106}}].

\bibitem{Gruppuso:2015xqa}
A.~Gruppuso, N.~Kitazawa, N.~Mandolesi, P.~Natoli and A.~Sagnotti,
  \emph{{Pre-Inflationary Relics in the CMB?}},
  \href{https://doi.org/10.1016/j.dark.2015.12.001}{\emph{Phys. Dark Univ.}
  {\bfseries 11} (2016) 68} [\href{https://arxiv.org/abs/1508.00411}{{\ttfamily
  1508.00411}}].

\bibitem{Hazra:coreforecast}
D.~K. Hazra, D.~Paoletti, M.~Ballardini, F.~Finelli, A.~Shafieloo, G.~F. Smoot
  et~al., \emph{{Probing features in inflaton potential and reionization
  history with future CMB space observations}},
  \href{https://doi.org/10.1088/1475-7516/2018/02/017}{\emph{JCAP} {\bfseries
  1802} (2018) 017} [\href{https://arxiv.org/abs/1710.01205}{{\ttfamily
  1710.01205}}].

\bibitem{Hergt:2018ksk}
L.~Hergt, W.~Handley, M.~Hobson and A.~Lasenby, \emph{{Constraining the
  kinetically dominated Universe}},
  \href{https://doi.org/10.1103/PhysRevD.100.023501}{\emph{Phys.\ Rev.\ D}
  {\bfseries 100} (2019) 023501}
  [\href{https://arxiv.org/abs/1809.07737}{{\ttfamily 1809.07737}}].

\bibitem{Ballardini:2018noo}
M.~Ballardini, \emph{{Probing primordial features with the primary CMB}},
  \href{https://doi.org/10.1016/j.dark.2018.11.006}{\emph{Phys. Dark Univ.}
  {\bfseries 23} (2019) 100245}
  [\href{https://arxiv.org/abs/1807.05521}{{\ttfamily 1807.05521}}].

\bibitem{Ragavendra:2019mek}
H.~Ragavendra, D.~Chowdhury and L.~Sriramkumar, \emph{{Unique contributions to
  the scalar bispectrum in `just enough inflation'}},
  \href{https://arxiv.org/abs/1906.03942}{{\ttfamily 1906.03942}}.

\bibitem{Ragavendra:2020old}
H.~Ragavendra, D.~Chowdhury and L.~Sriramkumar, \emph{{Suppression of scalar
  power on large scales and associated bispectra}},
  \href{https://arxiv.org/abs/2003.01099}{{\ttfamily 2003.01099}}.

\bibitem{Adams:2001vc}
J.~A. Adams, B.~Cresswell and R.~Easther, \emph{{Inflationary perturbations
  from a potential with a step}},
  \href{https://doi.org/10.1103/PhysRevD.64.123514}{\emph{Phys. Rev.}
  {\bfseries D64} (2001) 123514}
  [\href{https://arxiv.org/abs/astro-ph/0102236}{{\ttfamily
  astro-ph/0102236}}].

\bibitem{Chen:2006xjb}
X.~Chen, R.~Easther and E.~A. Lim, \emph{{Large Non-Gaussianities in Single
  Field Inflation}},
  \href{https://doi.org/10.1088/1475-7516/2007/06/023}{\emph{JCAP} {\bfseries
  06} (2007) 023} [\href{https://arxiv.org/abs/astro-ph/0611645}{{\ttfamily
  astro-ph/0611645}}].

\bibitem{Covi:2006ci}
L.~Covi, J.~Hamann, A.~Melchiorri, A.~Slosar and I.~Sorbera, \emph{{Inflation
  and WMAP three year data: Features have a Future!}},
  \href{https://doi.org/10.1103/PhysRevD.74.083509}{\emph{Phys. Rev.}
  {\bfseries D74} (2006) 083509}
  [\href{https://arxiv.org/abs/astro-ph/0606452}{{\ttfamily
  astro-ph/0606452}}].

\bibitem{Hazra:2010ve}
D.~K. Hazra, M.~Aich, R.~K. Jain, L.~Sriramkumar and T.~Souradeep,
  \emph{{Primordial features due to a step in the inflaton potential}},
  \href{https://doi.org/10.1088/1475-7516/2010/10/008}{\emph{JCAP} {\bfseries
  1010} (2010) 008} [\href{https://arxiv.org/abs/1005.2175}{{\ttfamily
  1005.2175}}].

\bibitem{Miranda:2012rm}
V.~Miranda, W.~Hu and P.~Adshead, \emph{{Warp Features in DBI Inflation}},
  \href{https://doi.org/10.1103/PhysRevD.86.063529}{\emph{Phys. Rev.}
  {\bfseries D86} (2012) 063529}
  [\href{https://arxiv.org/abs/1207.2186}{{\ttfamily 1207.2186}}].

\bibitem{Benetti:2013cja}
M.~Benetti, \emph{{Updating constraints on inflationary features in the
  primordial power spectrum with the Planck data}},
  \href{https://doi.org/10.1103/PhysRevD.88.087302}{\emph{Phys. Rev.}
  {\bfseries D88} (2013) 087302}
  [\href{https://arxiv.org/abs/1308.6406}{{\ttfamily 1308.6406}}].

\bibitem{Romano:2014kla}
A.~Gallego~Cadavid and A.~E. Romano, \emph{{Effects of discontinuities of the
  derivatives of the inflaton potential}},
  \href{https://doi.org/10.1140/epjc/s10052-015-3733-x}{\emph{Eur. Phys. J.}
  {\bfseries C75} (2015) 589}
  [\href{https://arxiv.org/abs/1404.2985}{{\ttfamily 1404.2985}}].

\bibitem{Chluba:2015bqa}
J.~Chluba, J.~Hamann and S.~P. Patil, \emph{{Features and New Physical Scales
  in Primordial Observables: Theory and Observation}},
  \href{https://doi.org/10.1142/S0218271815300232}{\emph{Int. J. Mod. Phys.}
  {\bfseries D24} (2015) 1530023}
  [\href{https://arxiv.org/abs/1505.01834}{{\ttfamily 1505.01834}}].

\bibitem{Ballardini:2017qwq}
M.~Ballardini, F.~Finelli, R.~Maartens and L.~Moscardini, \emph{{Probing
  primordial features with next-generation photometric and radio surveys}},
  \href{https://doi.org/10.1088/1475-7516/2018/04/044}{\emph{JCAP} {\bfseries
  1804} (2018) 044} [\href{https://arxiv.org/abs/1712.07425}{{\ttfamily
  1712.07425}}].

\bibitem{Martin:2003sg}
J.~Martin and C.~Ringeval, \emph{{Superimposed oscillations in the WMAP
  data?}}, \href{https://doi.org/10.1103/PhysRevD.69.083515}{\emph{Phys. Rev.}
  {\bfseries D69} (2004) 083515}
  [\href{https://arxiv.org/abs/astro-ph/0310382}{{\ttfamily
  astro-ph/0310382}}].

\bibitem{Ashoorioon:2006wc}
A.~Ashoorioon and A.~Krause, \emph{{Power Spectrum and Signatures for Cascade
  Inflation}},  \href{https://arxiv.org/abs/hep-th/0607001}{{\ttfamily
  hep-th/0607001}}.

\bibitem{Chen:2008wn}
X.~Chen, R.~Easther and E.~A. Lim, \emph{{Generation and Characterization of
  Large Non-Gaussianities in Single Field Inflation}},
  \href{https://doi.org/10.1088/1475-7516/2008/04/010}{\emph{JCAP} {\bfseries
  04} (2008) 010} [\href{https://arxiv.org/abs/0801.3295}{{\ttfamily
  0801.3295}}].

\bibitem{Biswas:2010si}
T.~Biswas, A.~Mazumdar and A.~Shafieloo, \emph{{Wiggles in the cosmic microwave
  background radiation: echoes from non-singular cyclic-inflation}},
  \href{https://doi.org/10.1103/PhysRevD.82.123517}{\emph{Phys. Rev.}
  {\bfseries D82} (2010) 123517}
  [\href{https://arxiv.org/abs/1003.3206}{{\ttfamily 1003.3206}}].

\bibitem{Flauger:2009ab}
R.~Flauger, L.~McAllister, E.~Pajer, A.~Westphal and G.~Xu, \emph{{Oscillations
  in the CMB from Axion Monodromy Inflation}},
  \href{https://doi.org/10.1088/1475-7516/2010/06/009}{\emph{JCAP} {\bfseries
  1006} (2010) 009} [\href{https://arxiv.org/abs/0907.2916}{{\ttfamily
  0907.2916}}].

\bibitem{Pahud:2008ae}
C.~Pahud, M.~Kamionkowski and A.~R. Liddle, \emph{{Oscillations in the inflaton
  potential?}}, \href{https://doi.org/10.1103/PhysRevD.79.083503}{\emph{Phys.
  Rev.} {\bfseries D79} (2009) 083503}
  [\href{https://arxiv.org/abs/0807.0322}{{\ttfamily 0807.0322}}].

\bibitem{Achucarro:2010jv}
A.~Achucarro, J.-O. Gong, S.~Hardeman, G.~A. Palma and S.~P. Patil, \emph{{Mass
  hierarchies and non-decoupling in multi-scalar field dynamics}},
  \href{https://doi.org/10.1103/PhysRevD.84.043502}{\emph{Phys. Rev.}
  {\bfseries D84} (2011) 043502}
  [\href{https://arxiv.org/abs/1005.3848}{{\ttfamily 1005.3848}}].

\bibitem{Aich:2011qv}
M.~Aich, D.~K. Hazra, L.~Sriramkumar and T.~Souradeep, \emph{{Oscillations in
  the inflaton potential: Complete numerical treatment and comparison with the
  recent and forthcoming CMB datasets}},
  \href{https://doi.org/10.1103/PhysRevD.87.083526}{\emph{Phys. Rev.}
  {\bfseries D87} (2013) 083526}
  [\href{https://arxiv.org/abs/1106.2798}{{\ttfamily 1106.2798}}].

\bibitem{Hazra:2012vs}
D.~K. Hazra, \emph{{Changes in the halo formation rates due to features in the
  primordial spectrum}},
  \href{https://doi.org/10.1088/1475-7516/2013/03/003}{\emph{JCAP} {\bfseries
  1303} (2013) 003} [\href{https://arxiv.org/abs/1210.7170}{{\ttfamily
  1210.7170}}].

\bibitem{Peiris:2013opa}
H.~Peiris, R.~Easther and R.~Flauger, \emph{{Constraining Monodromy
  Inflation}}, \href{https://doi.org/10.1088/1475-7516/2013/09/018}{\emph{JCAP}
  {\bfseries 1309} (2013) 018}
  [\href{https://arxiv.org/abs/1303.2616}{{\ttfamily 1303.2616}}].

\bibitem{Meerburg:2013dla}
P.~D. Meerburg and D.~N. Spergel, \emph{{Searching for oscillations in the
  primordial power spectrum. II. Constraints from Planck data}},
  \href{https://doi.org/10.1103/PhysRevD.89.063537}{\emph{Phys. Rev.}
  {\bfseries D89} (2014) 063537}
  [\href{https://arxiv.org/abs/1308.3705}{{\ttfamily 1308.3705}}].

\bibitem{Easther:2013kla}
R.~Easther and R.~Flauger, \emph{{Planck Constraints on Monodromy Inflation}},
  \href{https://doi.org/10.1088/1475-7516/2014/02/037}{\emph{JCAP} {\bfseries
  1402} (2014) 037} [\href{https://arxiv.org/abs/1308.3736}{{\ttfamily
  1308.3736}}].

\bibitem{Chen:2014cwa}
X.~Chen, M.~H. Namjoo and Y.~Wang, \emph{{Models of the Primordial Standard
  Clock}}, \href{https://doi.org/10.1088/1475-7516/2015/02/027}{\emph{JCAP}
  {\bfseries 1502} (2015) 027}
  [\href{https://arxiv.org/abs/1411.2349}{{\ttfamily 1411.2349}}].

\bibitem{Motohashi:2015hpa}
H.~Motohashi and W.~Hu, \emph{{Running from Features: Optimized Evaluation of
  Inflationary Power Spectra}},
  \href{https://doi.org/10.1103/PhysRevD.92.043501}{\emph{Phys. Rev.}
  {\bfseries D92} (2015) 043501}
  [\href{https://arxiv.org/abs/1503.04810}{{\ttfamily 1503.04810}}].

\bibitem{Miranda:2015cea}
V.~Miranda, W.~Hu, C.~He and H.~Motohashi, \emph{{Nonlinear Excitations in
  Inflationary Power Spectra}},
  \href{https://doi.org/10.1103/PhysRevD.93.023504}{\emph{Phys. Rev.}
  {\bfseries D93} (2016) 023504}
  [\href{https://arxiv.org/abs/1510.07580}{{\ttfamily 1510.07580}}].

\bibitem{Ballardini:2016hpi}
M.~Ballardini, F.~Finelli, C.~Fedeli and L.~Moscardini, \emph{{Probing
  primordial features with future galaxy surveys}},
  \href{https://doi.org/10.1088/1475-7516/2016/10/041}{\emph{JCAP} {\bfseries
  10} (2016) 041} [\href{https://arxiv.org/abs/1606.03747}{{\ttfamily
  1606.03747}}].

\bibitem{Ballardini:2019tuc}
M.~Ballardini, R.~Murgia, M.~Baldi, F.~Finelli and M.~Viel, \emph{{Non-linear
  damping of superimposed primordial oscillations on the matter power spectrum
  in galaxy surveys}},  \href{https://arxiv.org/abs/1912.12499}{{\ttfamily
  1912.12499}}.

\bibitem{Wands:2007bd}
D.~Wands, \emph{{Multiple field inflation}}, {\emph{Lect. Notes Phys.}
  {\bfseries 738} (2008) 275}
  [\href{https://arxiv.org/abs/astro-ph/0702187}{{\ttfamily
  astro-ph/0702187}}].

\bibitem{Kofman:1988xg}
L.~A. Kofman and D.~{\relax Yu}. Pogosian, \emph{{Nonflat Perturbations in
  Inflationary Cosmology}},
  \href{https://doi.org/10.1016/0370-2693(88)90109-8}{\emph{Phys. Lett.}
  {\bfseries B214} (1988) 508}.

\bibitem{Price:2014xpa}
L.~C. Price, J.~Frazer, J.~Xu, H.~V. Peiris and R.~Easther,
  \emph{{MultiModeCode: An efficient numerical solver for multifield
  inflation}}, \href{https://doi.org/10.1088/1475-7516/2015/03/005}{\emph{JCAP}
  {\bfseries 1503} (2015) 005}
  [\href{https://arxiv.org/abs/1410.0685}{{\ttfamily 1410.0685}}].

\bibitem{Chung:1999ve}
D.~J.~H. Chung, E.~W. Kolb, A.~Riotto and I.~I. Tkachev, \emph{{Probing
  Planckian physics: Resonant production of particles during inflation and
  features in the primordial power spectrum}},
  \href{https://doi.org/10.1103/PhysRevD.62.043508}{\emph{Phys. Rev.}
  {\bfseries D62} (2000) 043508}
  [\href{https://arxiv.org/abs/hep-ph/9910437}{{\ttfamily hep-ph/9910437}}].

\bibitem{Pi:2017gih}
S.~Pi, Y.-l. Zhang, Q.-G. Huang and M.~Sasaki, \emph{{Scalaron from
  $R^2$-gravity as a heavy field}},
  \href{https://doi.org/10.1088/1475-7516/2018/05/042}{\emph{JCAP} {\bfseries
  1805} (2018) 042} [\href{https://arxiv.org/abs/1712.09896}{{\ttfamily
  1712.09896}}].

\bibitem{Mori:2017caa}
T.~Mori, K.~Kohri and J.~White, \emph{{Multi-field effects in a simple
  extension of $R^2$ inflation}},
  \href{https://doi.org/10.1088/1475-7516/2017/10/044}{\emph{JCAP} {\bfseries
  1710} (2017) 044} [\href{https://arxiv.org/abs/1705.05638}{{\ttfamily
  1705.05638}}].

\bibitem{Gundhi:2018wyz}
A.~Gundhi and C.~F. Steinwachs, \emph{{Scalaron-Higgs inflation}},
  \href{https://doi.org/10.1016/j.nuclphysb.2020.114989}{\emph{Nucl. Phys.}
  {\bfseries B954} (2020) 114989}
  [\href{https://arxiv.org/abs/1810.10546}{{\ttfamily 1810.10546}}].

\bibitem{Gao:2012uq}
X.~Gao, D.~Langlois and S.~Mizuno, \emph{{Influence of heavy modes on
  perturbations in multiple field inflation}},
  \href{https://doi.org/10.1088/1475-7516/2012/10/040}{\emph{JCAP} {\bfseries
  1210} (2012) 040} [\href{https://arxiv.org/abs/1205.5275}{{\ttfamily
  1205.5275}}].

\bibitem{Achucarro:2010da}
A.~Achucarro, J.-O. Gong, S.~Hardeman, G.~A. Palma and S.~P. Patil,
  \emph{{Features of heavy physics in the CMB power spectrum}},
  \href{https://doi.org/10.1088/1475-7516/2011/01/030}{\emph{JCAP} {\bfseries
  1101} (2011) 030} [\href{https://arxiv.org/abs/1010.3693}{{\ttfamily
  1010.3693}}].

\bibitem{Polarski:1992dq}
D.~Polarski and A.~A. Starobinsky, \emph{{Spectra of perturbations produced by
  double inflation with an intermediate matter dominated stage}},
  \href{https://doi.org/10.1016/0550-3213(92)90062-G}{\emph{Nucl. Phys.}
  {\bfseries B385} (1992) 623}.

\bibitem{Feng:2003zua}
B.~Feng and X.~Zhang, \emph{{Double inflation and the low cmb quadrupole}},
  \href{https://doi.org/10.1016/j.physletb.2003.07.065}{\emph{Phys. Lett.}
  {\bfseries B570} (2003) 145}
  [\href{https://arxiv.org/abs/astro-ph/0305020}{{\ttfamily
  astro-ph/0305020}}].

\bibitem{Berkin:1991nm}
A.~L. Berkin and K.-I. Maeda, \emph{{Inflation in generalized Einstein
  theories}}, \href{https://doi.org/10.1103/PhysRevD.44.1691}{\emph{Phys. Rev.}
  {\bfseries D44} (1991) 1691}.

\bibitem{Starobinsky:1994mh}
A.~A. Starobinsky and J.~Yokoyama, \emph{{Density fluctuations in Brans-Dicke
  inflation}},  in \emph{{Proceedings, Workshop on General Relativity and
  Gravitation (JGRG4): Kyoto, Japan, November 28-December 1, 1994}}, p.~381,
  1994, \href{https://arxiv.org/abs/gr-qc/9502002}{{\ttfamily gr-qc/9502002}}.

\bibitem{GarciaBellido:1995fz}
J.~Garcia-Bellido and D.~Wands, \emph{{Constraints from inflation on scalar -
  tensor gravity theories}},
  \href{https://doi.org/10.1103/PhysRevD.52.6739}{\emph{Phys. Rev.} {\bfseries
  D52} (1995) 6739} [\href{https://arxiv.org/abs/gr-qc/9506050}{{\ttfamily
  gr-qc/9506050}}].

\bibitem{Starobinsky:2001xq}
A.~A. Starobinsky, S.~Tsujikawa and J.~Yokoyama, \emph{{Cosmological
  perturbations from multifield inflation in generalized Einstein theories}},
  \href{https://doi.org/10.1016/S0550-3213(01)00322-4}{\emph{Nucl. Phys.}
  {\bfseries B610} (2001) 383}
  [\href{https://arxiv.org/abs/astro-ph/0107555}{{\ttfamily
  astro-ph/0107555}}].

\bibitem{DiMarco:2002eb}
F.~Di~Marco, F.~Finelli and R.~Brandenberger, \emph{{Adiabatic and isocurvature
  perturbations for multifield generalized Einstein models}},
  \href{https://doi.org/10.1103/PhysRevD.67.063512}{\emph{Phys. Rev.}
  {\bfseries D67} (2003) 063512}
  [\href{https://arxiv.org/abs/astro-ph/0211276}{{\ttfamily
  astro-ph/0211276}}].

\bibitem{DiMarco:2005nq}
F.~Di~Marco and F.~Finelli, \emph{{Slow-roll inflation for generalized
  two-field Lagrangians}},
  \href{https://doi.org/10.1103/PhysRevD.71.123502}{\emph{Phys. Rev.}
  {\bfseries D71} (2005) 123502}
  [\href{https://arxiv.org/abs/astro-ph/0505198}{{\ttfamily
  astro-ph/0505198}}].

\bibitem{Lalak:2007vi}
Z.~Lalak, D.~Langlois, S.~Pokorski and K.~Turzynski, \emph{{Curvature and
  isocurvature perturbations in two-field inflation}},
  \href{https://doi.org/10.1088/1475-7516/2007/07/014}{\emph{JCAP} {\bfseries
  0707} (2007) 014} [\href{https://arxiv.org/abs/0704.0212}{{\ttfamily
  0704.0212}}].

\bibitem{Gordon:2000hv}
C.~Gordon, D.~Wands, B.~A. Bassett and R.~Maartens, \emph{{Adiabatic and
  entropy perturbations from inflation}},
  \href{https://doi.org/10.1103/PhysRevD.63.023506}{\emph{Phys. Rev.}
  {\bfseries D63} (2001) 023506}
  [\href{https://arxiv.org/abs/astro-ph/0009131}{{\ttfamily
  astro-ph/0009131}}].

\bibitem{GrootNibbelink:2001qt}
S.~Groot~Nibbelink and B.~J.~W. van Tent, \emph{{Scalar perturbations during
  multiple field slow-roll inflation}},
  \href{https://doi.org/10.1088/0264-9381/19/4/302}{\emph{Class. Quant. Grav.}
  {\bfseries 19} (2002) 613}
  [\href{https://arxiv.org/abs/hep-ph/0107272}{{\ttfamily hep-ph/0107272}}].

\bibitem{Schwarz:2001vv}
D.~J. Schwarz, C.~A. Terrero-Escalante and A.~A. Garcia, \emph{{Higher order
  corrections to primordial spectra from cosmological inflation}},
  \href{https://doi.org/10.1016/S0370-2693(01)01036-X}{\emph{Phys. Lett.}
  {\bfseries B517} (2001) 243}
  [\href{https://arxiv.org/abs/astro-ph/0106020}{{\ttfamily
  astro-ph/0106020}}].

\bibitem{Mukhanov:1990me}
V.~F. Mukhanov, H.~A. Feldman and R.~H. Brandenberger, \emph{{Theory of
  cosmological perturbations. Part 1. Classical perturbations. Part 2. Quantum
  theory of perturbations. Part 3. Extensions}},
  \href{https://doi.org/10.1016/0370-1573(92)90044-Z}{\emph{Phys. Rept.}
  {\bfseries 215} (1992) 203}.

\bibitem{Cremonini:2010ua}
S.~Cremonini, Z.~Lalak and K.~Turzynski, \emph{{Strongly Coupled Perturbations
  in Two-Field Inflationary Models}},
  \href{https://doi.org/10.1088/1475-7516/2011/03/016}{\emph{JCAP} {\bfseries
  1103} (2011) 016} [\href{https://arxiv.org/abs/1010.3021}{{\ttfamily
  1010.3021}}].

\bibitem{Tsujikawa:2002qx}
S.~Tsujikawa, D.~Parkinson and B.~A. Bassett, \emph{{Correlation - consistency
  cartography of the double inflation landscape}},
  \href{https://doi.org/10.1103/PhysRevD.67.083516}{\emph{Phys. Rev.}
  {\bfseries D67} (2003) 083516}
  [\href{https://arxiv.org/abs/astro-ph/0210322}{{\ttfamily
  astro-ph/0210322}}].

\bibitem{vandeBruck:2014ata}
C.~van~de Bruck and M.~Robinson, \emph{{Power Spectra beyond the Slow Roll
  Approximation in Theories with Non-Canonical Kinetic Terms}},
  \href{https://doi.org/10.1088/1475-7516/2014/08/024}{\emph{JCAP} {\bfseries
  1408} (2014) 024} [\href{https://arxiv.org/abs/1404.7806}{{\ttfamily
  1404.7806}}].

\bibitem{Silk:1986vc}
J.~Silk and M.~S. Turner, \emph{{Double Inflation}},
  \href{https://doi.org/10.1103/PhysRevD.35.419}{\emph{Phys. Rev.} {\bfseries
  D35} (1987) 419}.

\bibitem{Polarski:1994rz}
D.~Polarski and A.~A. Starobinsky, \emph{{Isocurvature perturbations in
  multiple inflationary models}},
  \href{https://doi.org/10.1103/PhysRevD.50.6123}{\emph{Phys. Rev.} {\bfseries
  D50} (1994) 6123} [\href{https://arxiv.org/abs/astro-ph/9404061}{{\ttfamily
  astro-ph/9404061}}].

\bibitem{Lesgourgues:1997ki}
J.~Lesgourgues and D.~Polarski, \emph{{CMB anisotropy predictions for a model
  of double inflation}},
  \href{https://doi.org/10.1103/PhysRevD.56.6425}{\emph{Phys. Rev.} {\bfseries
  D56} (1997) 6425} [\href{https://arxiv.org/abs/astro-ph/9710083}{{\ttfamily
  astro-ph/9710083}}].

\bibitem{Bond:2006nc}
J.~R. Bond, L.~Kofman, S.~Prokushkin and P.~M. Vaudrevange, \emph{{Roulette
  inflation with Kahler moduli and their axions}},
  \href{https://doi.org/10.1103/PhysRevD.75.123511}{\emph{Phys. Rev.}
  {\bfseries D75} (2007) 123511}
  [\href{https://arxiv.org/abs/hep-th/0612197}{{\ttfamily hep-th/0612197}}].

\bibitem{Kallosh:2018zsi}
R.~Kallosh, A.~Linde and Y.~Yamada, \emph{{Planck 2018 and Brane Inflation
  Revisited}}, \href{https://doi.org/10.1007/JHEP01(2019)008}{\emph{JHEP}
  {\bfseries 01} (2019) 008}
  [\href{https://arxiv.org/abs/1811.01023}{{\ttfamily 1811.01023}}].

\bibitem{Aghanim:2018eyx}
{\scshape Planck} collaboration, N.~Aghanim et~al., \emph{{Planck 2018 results.
  VI. Cosmological parameters}},
  \href{https://arxiv.org/abs/1807.06209}{{\ttfamily 1807.06209}}.

\bibitem{Braglia:2020eai}
M.~Braglia, D.~K. Hazra, F.~Finelli, G.~F. Smoot, L.~Sriramkumar and A.~A.
  Starobinsky, \emph{{Generating PBHs and small-scale GWs in two-field models
  of inflation}},
  \href{https://doi.org/10.1088/1475-7516/2020/08/001}{\emph{JCAP} {\bfseries
  08} (2020) 001} [\href{https://arxiv.org/abs/2005.02895}{{\ttfamily
  2005.02895}}].

\bibitem{Linde:2018hmx}
A.~Linde, D.-G. Wang, Y.~Welling, Y.~Yamada and A.~Achúcarro,
  \emph{{Hypernatural inflation}},
  \href{https://doi.org/10.1088/1475-7516/2018/07/035}{\emph{JCAP} {\bfseries
  07} (2018) 035} [\href{https://arxiv.org/abs/1803.09911}{{\ttfamily
  1803.09911}}].

\bibitem{Iarygina:2018kee}
O.~Iarygina, E.~I. Sfakianakis, D.-G. Wang and A.~Achucarro,
  \emph{{Universality and scaling in multi-field $\alpha$-attractor
  preheating}},
  \href{https://doi.org/10.1088/1475-7516/2019/06/027}{\emph{JCAP} {\bfseries
  06} (2019) 027} [\href{https://arxiv.org/abs/1810.02804}{{\ttfamily
  1810.02804}}].

\bibitem{Pi:2019ihn}
S.~Pi, M.~Sasaki and Y.-l. Zhang, \emph{{Primordial Tensor Perturbation in
  Double Inflationary Scenario with a Break}},
  \href{https://doi.org/10.1088/1475-7516/2019/06/049}{\emph{JCAP} {\bfseries
  06} (2019) 049} [\href{https://arxiv.org/abs/1904.06304}{{\ttfamily
  1904.06304}}].

\bibitem{Hazumi:2019lys}
M.~Hazumi et~al., \emph{{LiteBIRD: A Satellite for the Studies of B-Mode
  Polarization and Inflation from Cosmic Background Radiation Detection}},
  \href{https://doi.org/10.1007/s10909-019-02150-5}{\emph{J. Low. Temp. Phys.}
  {\bfseries 194} (2019) 443}.

\bibitem{Delabrouille:2017rct}
{\scshape CORE} collaboration, J.~Delabrouille et~al., \emph{{Exploring cosmic
  origins with CORE: Survey requirements and mission design}},
  \href{https://doi.org/10.1088/1475-7516/2018/04/014}{\emph{JCAP} {\bfseries
  1804} (2018) 014} [\href{https://arxiv.org/abs/1706.04516}{{\ttfamily
  1706.04516}}].

\bibitem{Hanany:2019wrm}
M.~Alvarez et~al., \emph{{PICO: Probe of Inflation and Cosmic Origins}},
  \href{https://arxiv.org/abs/1908.07495}{{\ttfamily 1908.07495}}.

\bibitem{Hanany:2019lle}
{\scshape NASA PICO} collaboration, S.~Hanany et~al., \emph{{PICO: Probe of
  Inflation and Cosmic Origins}},
  \href{https://arxiv.org/abs/1902.10541}{{\ttfamily 1902.10541}}.

\bibitem{Bingo}
D.~K. Hazra, L.~Sriramkumar and J.~Martin, \emph{Bingo: a code for the
  efficient computation of the scalar bi-spectrum}, {\emph{Journal of Cosmology
  and Astroparticle Physics} {\bfseries 2013} (2013) 026}.

\bibitem{Blas:2011rf}
D.~Blas, J.~Lesgourgues and T.~Tram, \emph{{The Cosmic Linear Anisotropy
  Solving System (CLASS) II: Approximation schemes}},
  \href{https://doi.org/10.1088/1475-7516/2011/07/034}{\emph{JCAP} {\bfseries
  1107} (2011) 034} [\href{https://arxiv.org/abs/1104.2933}{{\ttfamily
  1104.2933}}].

\bibitem{Bennett2013}
C.~L. Bennett, D.~Larson, J.~L. Weiland, N.~Jarosik, G.~Hinshaw, N.~Odegard
  et~al., \emph{Nine-yearwilkinson microwave anisotropy probe(wmap)
  observations: Final maps and results},
  \href{https://doi.org/10.1088/0067-0049/208/2/20}{\emph{The Astrophysical
  Journal Supplement Series} {\bfseries 208} (2013) 20}.

\bibitem{Linde:1983gd}
A.~D. Linde, \emph{{Chaotic Inflation}},
  \href{https://doi.org/10.1016/0370-2693(83)90837-7}{\emph{Phys. Lett.}
  {\bfseries 129B} (1983) 177}.

\bibitem{Bartolo:2001rt}
N.~Bartolo, S.~Matarrese and A.~Riotto, \emph{{Adiabatic and isocurvature
  perturbations from inflation: Power spectra and consistency relations}},
  \href{https://doi.org/10.1103/PhysRevD.64.123504}{\emph{Phys. Rev.}
  {\bfseries D64} (2001) 123504}
  [\href{https://arxiv.org/abs/astro-ph/0107502}{{\ttfamily
  astro-ph/0107502}}].

\bibitem{Wands:2002bn}
D.~Wands, N.~Bartolo, S.~Matarrese and A.~Riotto, \emph{{An Observational test
  of two-field inflation}},
  \href{https://doi.org/10.1103/PhysRevD.66.043520}{\emph{Phys. Rev.}
  {\bfseries D66} (2002) 043520}
  [\href{https://arxiv.org/abs/astro-ph/0205253}{{\ttfamily
  astro-ph/0205253}}].

\bibitem{Byrnes:2006fr}
C.~T. Byrnes and D.~Wands, \emph{{Curvature and isocurvature perturbations from
  two-field inflation in a slow-roll expansion}},
  \href{https://doi.org/10.1103/PhysRevD.74.043529}{\emph{Phys. Rev.}
  {\bfseries D74} (2006) 043529}
  [\href{https://arxiv.org/abs/astro-ph/0605679}{{\ttfamily
  astro-ph/0605679}}].

\bibitem{Dimopoulos:2005ac}
S.~Dimopoulos, S.~Kachru, J.~McGreevy and J.~G. Wacker, \emph{{N-flation}},
  \href{https://doi.org/10.1088/1475-7516/2008/08/003}{\emph{JCAP} {\bfseries
  0808} (2008) 003} [\href{https://arxiv.org/abs/hep-th/0507205}{{\ttfamily
  hep-th/0507205}}].

\bibitem{Easther:2013rva}
R.~Easther, J.~Frazer, H.~V. Peiris and L.~C. Price, \emph{{Simple predictions
  from multifield inflationary models}},
  \href{https://doi.org/10.1103/PhysRevLett.112.161302}{\emph{Phys. Rev. Lett.}
  {\bfseries 112} (2014) 161302}
  [\href{https://arxiv.org/abs/1312.4035}{{\ttfamily 1312.4035}}].

\bibitem{Price:2014ufa}
L.~C. Price, H.~V. Peiris, J.~Frazer and R.~Easther, \emph{{Gravitational wave
  consistency relations for multifield inflation}},
  \href{https://doi.org/10.1103/PhysRevLett.114.031301}{\emph{Phys. Rev. Lett.}
  {\bfseries 114} (2015) 031301}
  [\href{https://arxiv.org/abs/1409.2498}{{\ttfamily 1409.2498}}].

\bibitem{Achucarro:2015rfa}
A.~Achúcarro, V.~Atal and Y.~Welling, \emph{{On the viability of $m^2\phi^2$
  and natural inflation}},
  \href{https://doi.org/10.1088/1475-7516/2015/07/008}{\emph{JCAP} {\bfseries
  1507} (2015) 008} [\href{https://arxiv.org/abs/1503.07486}{{\ttfamily
  1503.07486}}].

\bibitem{Renaux-Petel:2015mga}
S.~Renaux-Petel and K.~Turzyński, \emph{{Geometrical Destabilization of
  Inflation}},
  \href{https://doi.org/10.1103/PhysRevLett.117.141301}{\emph{Phys. Rev. Lett.}
  {\bfseries 117} (2016) 141301}
  [\href{https://arxiv.org/abs/1510.01281}{{\ttfamily 1510.01281}}].

\bibitem{Langlois:2004nn}
D.~Langlois and F.~Vernizzi, \emph{{Mixed inflaton and curvaton
  perturbations}},
  \href{https://doi.org/10.1103/PhysRevD.70.063522}{\emph{Phys. Rev.}
  {\bfseries D70} (2004) 063522}
  [\href{https://arxiv.org/abs/astro-ph/0403258}{{\ttfamily
  astro-ph/0403258}}].

\bibitem{Moroi:2005kz}
T.~Moroi, T.~Takahashi and Y.~Toyoda, \emph{{Relaxing constraints on inflation
  models with curvaton}},
  \href{https://doi.org/10.1103/PhysRevD.72.023502}{\emph{Phys. Rev.}
  {\bfseries D72} (2005) 023502}
  [\href{https://arxiv.org/abs/hep-ph/0501007}{{\ttfamily hep-ph/0501007}}].

\bibitem{Moroi:2005np}
T.~Moroi and T.~Takahashi, \emph{{Implications of the curvaton on inflationary
  cosmology}}, \href{https://doi.org/10.1103/PhysRevD.72.023505}{\emph{Phys.
  Rev.} {\bfseries D72} (2005) 023505}
  [\href{https://arxiv.org/abs/astro-ph/0505339}{{\ttfamily
  astro-ph/0505339}}].

\bibitem{Ichikawa:2008iq}
K.~Ichikawa, T.~Suyama, T.~Takahashi and M.~Yamaguchi, \emph{{Non-Gaussianity,
  Spectral Index and Tensor Modes in Mixed Inflaton and Curvaton Models}},
  \href{https://doi.org/10.1103/PhysRevD.78.023513}{\emph{Phys. Rev.}
  {\bfseries D78} (2008) 023513}
  [\href{https://arxiv.org/abs/0802.4138}{{\ttfamily 0802.4138}}].

\bibitem{Enqvist:2013paa}
K.~Enqvist and T.~Takahashi, \emph{{Mixed Inflaton and Spectator Field Models
  after Planck}},
  \href{https://doi.org/10.1088/1475-7516/2013/10/034}{\emph{JCAP} {\bfseries
  1310} (2013) 034} [\href{https://arxiv.org/abs/1306.5958}{{\ttfamily
  1306.5958}}].

\bibitem{Vennin:2015vfa}
V.~Vennin, K.~Koyama and D.~Wands, \emph{{Encyclopædia curvatonis}},
  \href{https://doi.org/10.1088/1475-7516/2015/11/008}{\emph{JCAP} {\bfseries
  1511} (2015) 008} [\href{https://arxiv.org/abs/1507.07575}{{\ttfamily
  1507.07575}}].

\bibitem{Nicholson:2007by}
G.~Nicholson and C.~R. Contaldi, \emph{{The large scale CMB cut-off and the
  tensor-to-scalar ratio}},
  \href{https://doi.org/10.1088/1475-7516/2008/01/002}{\emph{JCAP} {\bfseries
  0801} (2008) 002} [\href{https://arxiv.org/abs/astro-ph/0701783}{{\ttfamily
  astro-ph/0701783}}].

\end{thebibliography}\endgroup
	%%%%%%%%%%%%%%%%%%%%%%%%%%%%%%%%%%%%%%%%%%%%%%%%%%%%%%%%%%%%%%%%%%%%%%%%%%%%%%%
\end{document}